\documentclass[journal]{IEEEtran}

\usepackage{wrapfig}
\usepackage{pgfplots}
\usepackage[center]{caption}
\usepackage{tikz}
\usepackage{tikzscale}
\usepackage{filecontents}    
\usepackage{amsmath}
\usepackage{mathtools}
\usepackage{amsfonts}
\usepackage{cite}
\usepackage[noend]{algorithmic}
\usepackage{algorithm}
\usepackage{array}
\usepackage{graphicx}
\usepackage{subcaption}
\usepackage{enumitem}
\usepackage{xcolor}
\newcommand{\fontsmall}{\fontsize{8pt}{9pt}\selectfont}
\usepackage{url}
\usepackage{citesort}

\begin{document}
\title{A Distributed Model-Free Ride-Sharing Approach for Joint Matching, Pricing, and Dispatching using Deep Reinforcement Learning}

\author{Marina~Haliem, 
	Ganapathy~Mani,
      Vaneet~Aggarwal, 
        and~Bharat~Bhargava
\thanks{MH, VA, and BB are with Purdue University, West Lafayette, IN, 47907 USA email: \{mwadea, bbshail,vaneet\}@purdue.edu. GM was with Purdue University when this work was completed, and is currently at Qualcomm, CA, email: manig@purdue.edu. This work was presented in part in ACM CSCS 2020 \cite{haliem2020distributed}. }
}
\maketitle

\begin{abstract}
Significant development of ride-sharing services presents a plethora of opportunities to transform urban mobility by providing personalized and convenient transportation while ensuring the efficiency of large-scale ride pooling. However, a core problem for such services is route planning for each driver to fulfill the dynamically arriving requests while satisfying given constraints. Current models are mostly limited to static routes with only two rides per vehicle (optimally) or three (with heuristics) \cite{alonso2017demand}, at least in the initial allocation while not ascertaining that opposite-direction rides are not grouped together.  In this paper, we present a dynamic, demand aware, and pricing-based vehicle-passenger matching and route planning framework that (1) dynamically generates optimal routes for each vehicle based on online demand, pricing associated with each ride, vehicle capacities and locations. This matching algorithm starts greedily and optimizes over time using an insertion operation, (2) involves drivers in the decision-making process by allowing them to propose a different price based on the expected reward for a particular ride as well as the destination locations for future rides, which is influenced by supply-and-demand computed by the Deep Q-network. (3) allows customers to accept or reject rides based on their set of preferences with respect to pricing and delay windows, vehicle type and carpooling preferences.  These (1-3) in tandem with each other enforce grouping rides with the most route-intersections together. (4) Based on demand prediction, our approach re-balances idle vehicles by dispatching them to the areas of anticipated high demand using deep Reinforcement Learning (RL). Our framework is validated using millions of trips extracted from the New York City Taxi public dataset; however, we consider different vehicle types and designed customer utility functions to validate the setup and study different settings. Experimental results show the effectiveness of our approach in real-time and large scale settings.
\end{abstract}


\IEEEpeerreviewmaketitle

\section{Introduction}
\subsection{Motivation}
Advanced user-centric ride-hailing services such as Uber and Lyft are thriving in urban environments by transforming urban mobility through convenience in travel to anywhere, by anyone, and at anytime. Given tens of millions of users per month \cite{2-2}, these mobility-on-demand (MoD) services introduce a new paradigm in urban mobility \textemdash ride-sharing or ride-splitting. These ride-hailing services, when adopting shared mobility, can provide an efficient and sustainable way of transportation \cite{36}. With higher usage per vehicle, its service can reduce traffic congestion, environmental pollution, as well as energy consumption, thereby enhance living conditions in urban environments \cite{40-37}. 
The growth in demand for these services coupled with the rise of self-driving technology points towards a need for a fleet management framework that accommodates both the drivers' and customers' preferences in an optimal and sustainable manner.  Even though the pooling services provide customized personal service to customers, both the drivers and the customers are largely left out in deciding what is best for them in terms of their conveniences and preferences. It is challenging to introduce customer and driver conveniences into the framework. For example, a customer may have a limitation on the money to spend on a particular ride as well as time constraints on reaching the destination. On the other hand, the driver may not be willing to accept the customer’s convenient fare as it may negatively affect his/her profits since the final destination may be in a low demand area. Thus,  a reliable framework is needed to identify trade-offs between drivers’ and customers’ needs and make a compromised decision that is favorable to both.  Our proposed intelligent transportation system is driven by the objective of maintaining service levels for customers while accomodating both drivers’ and customers’ preferences. Thus, it has the potential to revolutionize modern transportation systems.
\vspace{-.1in}
\subsection{Related Work}
\textbf{Dynamic Matching and Route-Planning:}
The assignment problem in urban transportation has been widely studied in literature. However, the majority of articles focus on static and deterministic Dial-A-Ride Problems (DARP)s with a homogeneous fleet of vehicles without pooling (e.g., \cite{ritzinger2016dynamic}); or on quantifying the system efficiency with a model-based approach \cite{zhan2016graph}.  But as concluded in \cite{HO2018395},  Dynamic and Stochastic DARPs appear to be the most challenging,  which is the focus of our paper. In \cite{agatz2012optimization},  the authors provide a review of optimization approaches in dynamic ride-sharing; whereas, the notion of shareability network is introduced in \cite{santi2014quantifying}. 
However, many of the approaches in literature limit their attention to variants of static ride-sharing, where it is assumed that all driver and rider requests are known prior to the execution of a matching process \cite{tafreshian2020trip,alonso2017demand}. However, these approaches are based on one-to-one matching and don't address dynamic route-planning where new ride requests can be served by a vehicle that might be already assigned a trip as long as it still has vacant seats.\\
In recent years,  dynamic ride-sharing started to gain attention (e.g, \cite{wang2018stable}).  Some articles deal with this planning uncertainty by using a rolling horizon solution approach \cite{AGATZ20111450,liang2020automated}. These approaches are not always applicable in real-time due to the need for frequent re-optimizations. To minimize this frequncy,  the authors in \cite{kleiner2011mechanism} commit the arrangement as late as possible given the time considerations.  However, neither drivers' nor customers' preferences are accommodated in the decision-making process.  
In \cite{zhang2020pricing},  the authors model the allocation problem as bipartite matching and address the drivers' scheduling problem by proposing a heuristic nearest neighbor algorithm. The algorithm in \cite{alonso2017demand} generates optimal routes using constrained optimization.  
However, both are still one-to-one approaches that do not take into consideration which rides to group together.  Some approaches consider Tabu search and heuristic algorithms for a robust optimization \cite{li2020ride}. However, RL is proven to reach near-optimal solutions that outperform metaheuristic algorithms \cite{HU2020106244}, especially on large scale problems like ours.  To further improve the matching,  \cite{liu2020mobility} proposes mT-share that fully exploits both real-time and historical mobility information of taxis and ride requests. \\
An essential issue in realizing shared mobility in online ride-sharing settings is \textit{route planning}, which has been proven to be an NP-hard problem \cite{6,DARP}. Route planning - given a set of vehicles $V$, and requests $R$ - designs a route that consists of a sequence of pickup and drop-off locations of requests assigned to each vehicle. In ride-sharing environments, vehicles and requests arrive dynamically and are arranged such that they meet different objectives (e.g., maximizing the fleet's profits \cite{10-DA,40-37}). 
Utilizing an operation called \textit{insertion} to solve such a highly dynamic problem has been proven, in literature, to be both effective and efficient \cite{12,14,32,38,40-37,li2020ride}. The \textit{insertion} operation aims to accommodate the newly arriving request by inserting its pickup (origin) and its drop-off (destination) locations into a vehicle’s route. 
Most of the approximation algorithms that provide solutions to the route planning problem are limited to only two requests per vehicle \cite{6,DARP}. In \cite{DARP}, the authors improve the effectiveness of route planning for shared mobility by considering the near-future demand, 
thus overcoming the short-sightedness problem of the insertion operator. However, their approach is still limited to only two rides per vehicle. Most approaches in literature don't enforce constraints on which rides can be grouped together, which may result in grouping rides whose destinations are in opposite directions; thus resulting in large detours (boosting the travel distance as well as fuel and energy consumption per vehicle). Therefore, a robust framework is essential to achieve a dynamic and demand-aware matching framework that optimizes for the overall fleet's objectives.  Finally, the overall problem for ride-sharing requires interconnected decisions of matching, pricing, and dispatching, and is the focus of this paper. \\
In this paper, we present a dynamic, demand-aware, and pricing-based vehicle-passenger matching and route-planning framework that scales up to accommodate more than two rides per vehicle (up to the maximum capacity of the vehicle) in the initial assignment phase followed by an optimization phase. 
	  Thus, our matching satisfies the platform-related set of constraints (described in Section \ref{darm}).  By incorporating a pricing strategy into our framework, we enforce grouping together rides with significant path-intersections.

\textbf{Joint Pricing and Matching:}
Regarding the pricing problem, there are various ways to divide the trip costs among the rideshare partners \cite{AGATZ20111450,WANG2019122,zhang2020pricing}. 
However, none of the works in literature combine pricing with matching, they study them as two separate problems. However, grouping requests together greatly affects the pricing decisions and vice-versa. Ascertaining grouping together requests that are not going in opposite directions has not been addressed in literature.  Most works in literature assume that vehicles are told which passengers go together, by taking near-by pickup locations together (e.g., \cite{xu2020efficient,deep_pool}) and thus are at the risk of grouping rides going in opposite directions. 
In this paper, we approach the pricing-based ride-sharing problem through a model-free technique for ride-sharing with ride-pooling. In contrast to the model-based approaches in literature \cite{3,4,kleiner2011mechanism,6}, our proposed approach can adapt to dynamic distributions of customer and driver preferences.  At the same time, we incorporate the pricing aspect to our matching algorithm which would prioritize rides with significant path-intersections to be grouped together. This provides an efficient dynamic pricing-aware matching algorithm for the ridesharing environment by managing the prices for passengers based on the distance traveled due to serving this passenger (i.e., inserting this request in the vehicle's current route). The intuition is that the insertion cost of a request leading to an opposite direction to any of the previously inserted requests will be very high, and thus will be highly unlikely to be inserted into this specific route.  If it were the only option, this high cost will cause the driver to propose a high increase in price, which will be highly likely rejected by the passenger. The only case where the passenger may accept this ride is if they are willing to tolerate both the extra cost and time delay,
which would mean the customer will end up satisfied with the service whereas the driver will be compensated for the extra fuel cost incurred by the extra pay. In this manner, the matching, route planning, and pricing components work in tandem with each other to reach a common ground solution that is favorable by and profitable to both parties.  In \cite{deep_pool}, the authors provided the first model-free approach for ride-sharing with pooling, DeepPool, using deep RL. However, DeepPool neither incorporates dynamic demand-aware matching with a pricing strategy, nor accommodates customers' and drivers' conveniences. It primarily focuses on dispatching idle vehicles. 
To the best of our knowledge, ours is the first work that introduces a model-free approach for a distributed joint matching, pricing, and dispatching where customers and drivers can weigh in their ride preferences, influencing the decision making of ride-sharing platforms.\\
\textbf{Ridesharing Dispatch:}
We note that the vehicle dispatching problem \cite{agatz2012optimization,deep_pool,fleet_oda} has also been referred to as repositioning \cite {alonso2017demand,ma2019dynamic} or rebalancing \cite{alonso2017demand}, of vehicles, in literature. 
Many of the studies on optimal taxi dispatching address the problem as either a variant of the vehicle routing problem (VRP) \cite{jung2016dynamic} or the bipartite graph matching problem \cite{agatz2012optimization,zhan2016graph,tafreshian2020trip}. Under the VRP formulation, each taxi is assigned to sequentially pick up a number of passengers and under the bipartite graph formulation, each taxi is matched with the closest passenger in its vicinity.  
However, neither of these approaches consider which ride requests to group together, they can potentially end up assigning requests with opposite-direction-destinations to the same vehicle.  To handle dynamic traffic conditions, \cite{alisoltani2020multi} utilizes the observed average speed and updates every 10 seconds, while \cite{ramezani2017dynamic} builds on the macroscopic fundamental diagram (MFD) concept. In \cite{vazifeh2018addressing}, authors provide a network-based solution to determine the minimum number of vehicles needed to serve all the trips. 
However,  they do not account for any other drivers' or passengers' preferences. 
In this paper, we devise a DQN reward function that accounts for passenger-specific as well as driver-specific objectives. We utilize the dispatch of idle vehicles using a Deep Q-Network (DQN) framework as in \cite{deep_pool}. 
The goal of our approach is to influence the customer and vehicle utility functions, utilizing an optimal dispatching framework, to (i) achieve convenient pricing and matching decisions, (ii) optimize vehicles' route planning (iii) re-position idle vehicles to areas of predicted high demand. 
Depending on the DQN, the pricing decisions are made where drivers can propose an additional price if the passenger's destination location is that of a predicted low demand (i.e., future reward) as explained in Section \ref{propose}.  Further, the DQN takes into account the profits of the vehicles (as part of its objective function) and thus is affected by both the pricing and matching decisions. Therefore,  the three components of pricing, matching, and dispatching influence each other, and are tightly coupled.

\vspace{-.1in}
\subsection{Contributions}
The key contributions, in this paper, can be summarized as: 
\begin{itemize}[leftmargin = *]
	\item We present a novel dynamic, Demand-Aware and Pricing-based Matching and route planning (DARM) framework that is scalable up to the maximum capacity per vehicle in the initial assignment phase. In the optimization phase, this algorithm takes into account the near-future demand as well as the pricing associated with each ride in order to improve the route-planning by eliminating rides heading towards opposite directions and applying insertion operations to vehicles' current routes. 
	\item In addition to our matching and route-planning (DARM) framework, we integrate a novel Distributed Pricing approach for Ride-Sharing -with pooling- (DPRS) framework where, based on their convenience, customers and drivers get to weigh-in on the decision-making of a particular ride. The key idea is that the passengers are offered price based on the additional distance given the previous matched passengers thus prioritizing the passengers which will have intersections in the routes to be grouped together. 
	\item In the DPRS framework, drivers are allowed to propose a price based on the location of the ride that accounts for the reward of DQN based on the destination location. Similarly, customers can either accept or reject rides based on their pricing and timing thresholds, vehicle type, and number of people to share a ride with. 
	\item Our joint (DARM + DPRS) framework increases the profit margins of both customers and drivers,  and the profits are also fed back to the reinforcement learning utility functions that  influence the Q-values learnt using DQN for making the vehicles' dispatch decisions. The optimization problem is formulated such that our novelty framework minimizes the rejection rate, customers' waiting time, vehicles' idle time, the total number of vehicles to reduce traffic congestion,  fuel consumption,  and maximizes the vehicle's profit.
	\item  We simulate the ride-sharing system\footnote{The code for this work is available at \url{https://github.itap.purdue.edu/Clan-labs/Dynamic_Matching_RS}. } using real-world dataset of New York City's taxi trip records (15 million trips)\cite{10}.  Experimental results show that our novel Joint (DARM + DPRS) framework provides $10$ times more profits for drivers when compared to various baselines, while maintaining waiting times of ${\small < 1}$ min. for customers. Besides, we show a significant improvement in fleet utilization, utilizing only ${\small 50\%}$ of the vehicles allowed by the system to fulfill ${\small \approx 96\%}$ of the demand,in contrast to baselines that utilize $> 80\%$ of allowed vehicles to serve ${\small <60\%}$ of the demand. 
\end{itemize}

The rest of this paper is organized as follows: Section \ref{joint} describes the overall architecture of our framework as well as the model parameters. Section \ref{darm},  explains our dynamic, demand-aware,  and pricing-based matching and route planning. In Section \ref{pricing}, we provide details for our pricing strategy, including customers' and drivers' utility functions and their decision-making processes.  In Section \ref{dqn_algo}, we describe the DQN-based approach utilized for dispatching idle vehicles. Simulation setup and experimental results are presented in Section \ref{results}. Finally, Section \ref{conc} concludes the paper.


\vspace{-0.05in}
\section{Distributed Joint Matching, Pricing and Dispatching Framework}\label{joint}
We propose a novel distributed framework for matching, pricing, and dispatching in ride-sharing environments using Deep Q-Network (DQN), where initial matchings (that are decided in a greedy fashion) are then optimized in a distributed manner (per vehicle) in order to meet the vehicle's capacity constraints as well as minimize customers' extra waiting time and driver's additional travel distance. This framework involves customers and drivers (will be referred to as Agents henceforth) in the decision-making process. They learn the best pricing actions based on their utility functions that dynamically change based on each agent's set of preferences and environmental variables. Moreover, vehicles learn the best future dispatch action to take at time step $t$, taking into consideration the locations of all other nearby vehicles, but without anticipating their future decisions. Note that, vehicles get dispatched to areas of anticipated high-demand either when they first enter the market, or when they spend a long time being idle (searching for a ride). Vehicles' dispatch decisions are made in parallel, since drivers learn the location updates of other vehicles in real-time (e.g., GPS), 
so it is unlikely for two drivers to take actions at the same exact time.  Therefore, our algorithm learns the optimal policy for each agent independently as opposed to centralized-based approaches \cite{fleet_oda}. 

\begin{figure}[th]
\captionsetup{justification=centering, font=small, format=hang}
\centering
\includegraphics[width=.48\textwidth]{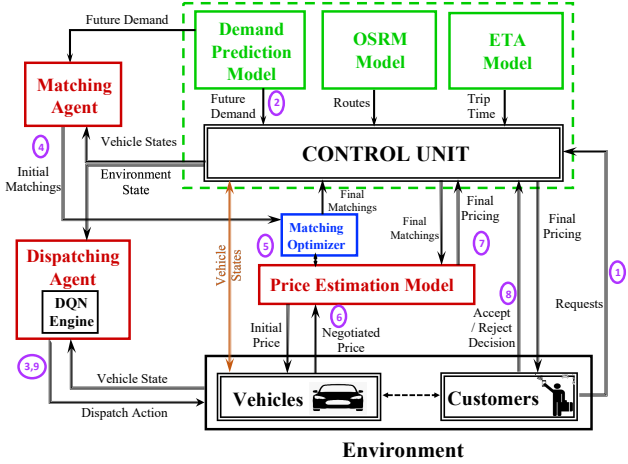}
\vspace{-0.8em}
\caption{\small Overall architecture of the proposed framework \protect \footnotemark} \label{arch}
\vspace{-1.8em}
\end{figure}

\vspace{-0.15in}
\subsection{Model Architecture}
\footnotetext{Enlarged figure is provided in Fig.  \ref{large} in Appendix D.}
Figure \ref{arch} shows the basic components of our joint framework and the interaction steps (in purple) between them. We assume that the central control unit is responsible for: (1) maintaining the states such as current locations, current capacity, destinations, etc., for all vehicles. These states are updated in every time step based on the dispatching and matching decisions. (2) The control unit also has some internal components that help manage the ride-sharing environment such as: (a) the estimated time of arrival (ETA) model used to calculate and continuously update the estimated arrival time. (b) The Open Source Routing Machine (OSRM) model used to generate the vehicle's optimal trajectory to reach a destination, and (c) the (Demand Prediction) model used to calculate the future anticipated demand in all zones. We adopt these three models from \cite{fleet_oda}; whose details and their relevant calculations are provided in Appendix \ref{models}. For every time step, first, the ride requests are input to the system (Step 1 in Fig. \ref{arch}) along with the heat map for supply and demand (which involves demand prediction in the near future) [Step 2 in Fig. \ref{arch}]. Then, based on the predicted demand, vehicles adopt a dispatching policy using DQN, where they get dispatched to zones with anticipated high demand (Step 3). This step not only takes place when vehicles first enter the market (lines 3-4 in Algorithm \ref{sim}), but also when they experience large idle durations (at the end of every time step, this gets checked for - lines 23-24 (Step 9)). Then, each vehicle receives the updated environment state from the control unit and performs initial greedy vehicle-passenger(s) matching (Step 4 in Fig.  \ref{arch}), where one request (or more) gets assigned to the nearest vehicle based on its maximum passenger capacity. Next, communicating with the Price Estimation model, each vehicle calculates the corresponding initial pricing associated with each request (Step 5).  
Afterward, 
each vehicle executes its matching optimizer module that performs an insertion-based route planning (Step 6).  In this step, vehicles reach their final matchings list by dealing with their initial matchings list in the order of their proximity, performing an insertion operation to its current route plan (as long as this insertion satisfies the capacity, extra waiting time, and additional travel distance constraints to guarantee that serving this request would yield a profit). Using the expected discounted reward learnt from DQN (in step 3), and the ride's destination, vehicles weigh their utility based on the potential hotspot locations, and propose new pricing for the customer (at the end of steps 5 and 6). This takes place on a customer-by-customer basis, where a vehicle upon inserting a customer into its current route plan, proposes to him/her the new price (Step 7). Then, the customer has the ability to accept or reject based on their own independent utility function (Step 8 in Fig. \ref{arch}). Finally, upon receiving the customer's decision, the driver either confirms the addition of this ride to its route plan or removes it.  Algorithm \ref{sim} shows the overall flow of our framework. Note that lines (7-22) that correspond to steps (3-9) take place in parallel at each vehicle. 
\\
The proposed model is distributed in the sense that each vehicle solves its own DQN and utilizes the output Q-values to make matching, pricing, route planning, and dispatching decisions, without communicating with other vehicles in its vicinity. However, a vehicle would consider the locations of nearby vehicles while making its independent decisions. So, a vehicle can communicate with the control unit, as needed, to request new information of the environment (prior to making decisions) or update its own status (after any decision).

\vspace{-.15in}
\subsection{Model Parameters and Notations}\label{MP}
We built a ride-sharing simulator to train and evaluate our framework. We simulate New York City as our area of operation, where the map is divided into multiple non-overlapping regions, a grid with each 1 square mile being taken as a zone. This allows us to discretize the area of operation and thus makes the action space\textemdash where to dispatch the vehicles\textemdash tractable. This discretization prevents our state and action space from exploding thereby making implementation feasible. We use {\small $m \in \{1, 2, 3, \cdots, M\}$} to denote the city's zones, and $n$ to denote the number of vehicles. A vehicle is marked as \textit{available} if it has any remaining seating capacity. 
Available vehicles in zone $i$ at time slot $t$ is denoted $v_{t,i}$. 
We optimize our algorithm over $T$ time steps, each of duration $\Delta t$. Idle vehicles make decisions on where on the map to head-to to serve the demand at each time step {\small $\tau = t_{0},t_{0}+\Delta t,t_{0}+2(\Delta t),\ldots,t_{0}+T(\Delta t)$} where $t_{0}$ is the start time. Below, we present our model parameters and notations:
\begin{enumerate}[leftmargin=*]
\item \textit{Demand:} We denote the number of requests for zone $m$ at time $t$ as $d_{t,m}$. The future pick-up request demand in each zone is predicted through a historical distribution of trips across the zones \cite{plan_article}, and is denoted by {\small $D_{t:T} = (\boldsymbol{\overline{d}_{t}},\ldots,\boldsymbol{\overline{d}_{t+T}})$ from time $t_{0}$ to $t + T$}.
\item \textit{Vehicle States:} We use {\small $X_{t} = \{x_{t,1}, x_{t,2}, \; ... \; , x_{t,N}\}$} to denote the $N$ vehicles' status at time $t$. $x_{t,n}$ tracks vehicle $n$'s state variables at time step $t$. For a given vehicle, we keep track of its: (1) current location/zone $V_{loc}$,  (2) current capacity $V_{C}$, (3) type $V_{T}$, (4) maximum capacity {\small $C_{max}^{V}$}, (5) Pick-up time for each passenger, and (6) the destination of each passenger.  A vehicle is considered available if at least one of its seats is vacant that is, if and only if $V_{C} < C_{max}^{V}$. 
\item \textit{Supply:} At each time slot $t$, the supply of vehicles for each zone is projected to future time $\tilde{t}$. $d_{t,\tilde{t}, m}$ is the number of vehicles that are currently unavailable at time $t$ but will become available at time $\tilde{t}$ as they will drop-off customer(s) at region $m$. This information can be ascertained using the ETA prediction for all vehicles. Consequently, 
we can predict the number of vehicles in each zone, from time $t_{0}$ to time $t+T$, denoted by $V_{t:t+T}$ which serves as our predicted supply in each zone for $T$ time slots ahead. 
\end{enumerate}
Our framework keeps track of the rapid changes of all these variables and seeks to make the demand, $d_{t}$, $\forall t$ and supply $v_{t}$, $\forall t$ close enough (mismatch between them is zero).

\begin{algorithm}[ht]
	\caption{ \small Joint RideSharing Framework} \label {sim}
		\fontsmall
		\begin{algorithmic}[1]
			\STATE \textbf{Initialize} vehicles' states $X_{0}$ at $t_{0}$.
			\FOR {$t \in T$}
				\STATE \textbf{Fetch} all ride requests at time slot t, $D_{t}$.
				\STATE \textbf{Fetch} all vehicles that entered the market in time slot t, $V_{new}$.
				\STATE \textbf{Dispatch} $V_{new}$ to zones with anticipated high demand - Algorithm \ref{dqn_brief}
				\STATE \textbf{Fetch} all available vehicles at time slot t, $V_{t}$.	
				\FOR {each vehicle $V_{j} \in V_{t} \ldots$}
					\STATE \textbf{Obtain} initial matching $A_{j}$ using Algorithm \ref{assign} in \ref{phase1}.
					\FOR {each ride request $r_{i} \in A_{j} \ldots$}
						\STATE \textbf{Obtain} initial price $P_{init}(r_{i})$ using \eqref{init}.
						\STATE \textbf{Perform} route planning using Algorithm \ref{insert} in \ref{phase2}.
						\STATE \textbf{Obtain} $ S^{\prime}_{V_{j}}[r_{i}]$ based on $\text{cost(} V_{j}, S^{\prime}_{V_{j}}[r_{i}])$.
						\STATE \textbf{Update} trip time $T_{i}$ based on $S^{\prime}_{V_{j}}[r_{i}]$ using ETA model.
						\STATE \textbf{Calculate} final price $P(r_{i})$ based on $S^{\prime}_{V_{j}}[r_{i}]$ using \eqref{car}.
						\STATE \textbf{Get} customer $i$'s decision $C^{i}_{d}$ on $P(r_{i})$ using \eqref{customer1} \& \eqref{customer2}.
						\IF {$C^{i}_{d} == 1$}
							\STATE \textbf{Update} $S_{V_{j}} \gets S^{\prime}_{V_{j}}[r_{i}]$.
						\ELSE
							\STATE \textbf{Insert} $r_{i}$ to $D_{t+1}$
						\ENDIF
					\STATE \textbf{Update} the state vector $s_{t}$.
					\ENDFOR
					\STATE \textbf{Retrieve} next stop from $S_{V_{j}}$.
					\STATE \textbf{Head} to next stop (whether a pickup or a dropoff).
				\ENDFOR
				\STATE \textbf{Fetch} all idle vehicles with $\text{Idle\_duration} > 10$ minutes, $V_{idle}$.
				\STATE \textbf{Dispatch} $V_{idle}$ to zones with anticipated high demand - Algorithm \ref{dqn_brief}
				\STATE \textbf{Update} the state vector $s_{t}$.
			\ENDFOR
		
		\end{algorithmic} 
\end{algorithm}

\vspace{-0.1in}
\section{DARM framework for Matching and Route Planning}\label{darm}
\textbf{NP-Hardness:}  The ride-sharing assignment problem is proven to be NP-hard in \cite{6} as it is a reduction from the 3-dimensional perfect matching problem (3DM). In 3DM, given a number of requests with source and destination locations and a number of available vehicle locations, the task is to assign vehicles to requests. However, in \cite{6}, the authors limit this allocation to only two requests sharing the same vehicle at a time using an approximation algorithm that is 2.5 times the optimal cost. They approach this problem by pairing requests first greedily, and then using bipartite graphs to match each vehicle to one pair of requests, while assigning the maximum number of requests with the minimum total cost. In our approach, we don't limit matching to only two requests; instead, we go as far as the maximum capacity of a vehicle allows (satisfying the capacity constraint), which significantly boosts the acceptance rate of passengers.\\
Our dynamic, demand-aware, and pricing-based DARM framework goes through two phases:

\vspace{-0.8em}
\subsection{Initial Vehicle-Passenger(s) Assignment Phase:} \label{phase1}
The initial assignment is represented in Algorithm \ref{assign}. In this phase, each vehicle having the knowledge of the future demand (fed from the control unit) $D_{t:t+T}$ at each zone, the vehicles' status vectors $X_{t}$ including their current locations as well as the origin $o_{i}$ and destination $d_{i}$ locations for each request $r_{i}$, performs a greedy matching operation. This is where each request $r_{i}$ gets assigned to the nearest available vehicle to it, satisfying the capacity constraints. In other words, we define the capacity constraint to be: the number of all requests assigned to vehicle $V_{j}$ is less than its maximum capacity $C^{V_{j}}_{max}$ at any time. At the end of this phase, each vehicle $V_{j}$ has a list of initial matchings $A_{j} = [r_{1}, r_{2}, ..., r_{k}]$, where $k \leq C^{V_{j}}_{max}$ assuming that each request has only one passenger. Assume the passenger count per request is $\mid r_{i} \mid $, and the vehicle $V_{j}$ arrives at location $z$. Then, to check the capacity constraint in $O(1)$ time, we define vehicle $V_{j}$'s current capacity $V^{j}_{C}[z]$  that refers to the total capacity of the requests that are still on-board of $V_{j}$ when it arrives at that location $z$ as follows:
\begin{equation}
{\small
V^{j}_{C}[z]  = \\
\begin{cases}
V^{j}_{C}[z-1] + \mid r_{i} \mid  & \text{if $z == o_{i}$} \\
V^{j}_{C}[z-1] - \mid r_{i} \mid  &  \text{if $z == d_{i}$} 
\end{cases}}
\end{equation}

\begin{algorithm}[!htb]
	\caption{Greedy Assignment} \label {assign}
	\begin{minipage}{.9\linewidth}
		\fontsmall
		\begin{algorithmic}[1]
		\STATE \textbf{Input: } Available Vehicles $V_{t}$ with their locations $loc(V_{j})$ such that: $V_{j} \in V_{t}$, Ride Requests $D_{t}$ with origin $o_{i}$ and destination $d_{i}$ associated with $r_{i} \in D_{t}$.
		\STATE \textbf{Output: } Matching decisions $A_{j}$ for each $V_{j} \in V_{t}$
		\STATE \textbf{Initialize} $A_{j} = [\;], V^{j}_{\text{capacity}} = V^{j}_{C}$ for each $V_{j} \in V_{t}$.
		\FOR {each $r_{i} \in D_{t} \dots$}
			\STATE \textbf{Obtain} locations of candidate vehicles $V_{\text{cand}}$, such that:\\ $|loc(V_{j}) - o_{i}| \leq 5 \; km^{2}$ \textbf{AND} $(V^{j}_{\text{capacity}} + |r_{i}|) \leq C^{V_{j}}_{max}$.
			\STATE \textbf{Calculate} trip time $T_{j,i} \in T_{\text{cand}, i}$ from each $loc(V_{j}) \in V_{\text{cand}}$ to $o_{i}$ using the ETA model.
			\STATE \textbf{Pick} $V_{j}$ whose $T_{j,i} = \text{argmin}(T_{\text{cand}, i})$ to serve ride $r_{i}$.
			\STATE \textbf{Push} $r_{i}$ to $A_{j}$
			\STATE \textbf{Update} $loc(V_{j}) \gets o_{i}$
			\STATE \textbf{Increment} $V^{j}_{\text{capacity}} \gets V^{j}_{\text{capacity}} + |r_{i}|$
		\ENDFOR
		\STATE \textbf{Return} $A_{t} = [A_{j}, A_{j+1}, ..., A_{n}]$, where $n = |V_{t}|$.
		\end{algorithmic} 
	\end{minipage}
\end{algorithm}

\vspace{-1em}
\subsection{Distributed Optimization Phase:}\label{phase2}
Our demand-aware route planning problem is a variation of the basic route planning problem (which is NP-hard) for shareable mobility services \cite{30} \cite{40-37} by setting $\alpha$ = 1 and $\beta = 0$.  Further, the existing literature proved that there is no optimal method to maximize the total revenue for the basic route planning problem (which is reducible to our DARM problem) using neither deterministic nor randomized algorithms \cite{10-DA} \cite{40-37}. Thus, the same applies to our DARM problem.  However, several studies show that Insertion is an effective approach to greedily deal with the shared mobility problem. We propose an insertion based framework,  similar to the idea in \cite{xu2020efficient},  to optimize our matching framework.  However, in \cite{xu2020efficient}, authors group together close-by requests without considering if their destinations are in opposite directions, while our approach mitigate this problem. Also,  \cite{xu2020efficient} use denwick tree to speed up their operations while we utilize our OSRM module as will be explained below. Note that, in our framework, this optimization step is impacted by the pricing decisions that are made based on the Q-values learnt from our DQN. In this aspect, our approach can achieve better results in a relatively long time period using the near future predicted demand (that is part of the DQN input) to overcome the short-sightedness problem of the basic insertion algorithms.

In DARM, we follow the idea of searching each route and locally optimally inserting new vertex (or vertices) into a route. In our problem, there are two vertices (i.e., origin $o_{i}$ and destination $d_{i}$) to be inserted for each request $r_{i}$. We define the insertion operation as: given a vehicle $V_{j}$ with the current route $S_{V_{j}}$,  and a new request $r_{i}$, the insertion operation aims to find a new feasible route $S^{\prime}_{V_{j}}$ by inserting $o_{i}$ and $d_{i}$ into $S_{V_{j}}$ with the minimum increased cost, that is the minimum extra travel distance, while maintaining the order of vertices in $S_{V_{j}}$ unchanged in $S^{\prime}_{V_{j}}$. Specifically, for a new request $r_{i}$, the basic insertion algorithm checks every possible position to insert the origin and destination locations and return the new route such that the incremental cost is minimized. To present our cost function, we first define our distance metric, where given a graph $G$ we use our OSRM engine to pre-calculate all possible routes over our simulated city. Then, we derive the distances of the trajectories (i.e., paths) from location $a$ to location $b$ to define our graph weights. Thus, we obtain a weighted graph $G$ with realisitic distance measures serving as its weights. We extend the weight notation to paths as follows:
 $w(a_{1}, a_{2}, ..., a_{n}) = \sum_{i=1}^{n-1} w(a_{i}, a_{i+1})$. \\
Thus, we define the cost associated with each new potential route/path $S^{\prime}_{V_{j}} = [r_{i}, r_{i+1}, ..., r_{k}]$ to be the $\text{cost}(V_{j}, S^{\prime}_{V_{j}}) = w(r_{i}, r_{i+1}, ... r_{k})$ resulting from this specific ordering of vertices (origin and destination locations of the $k$ requests assigned to vehicle $V_{j}$). Besides, we derive the cost of the original route to calculate the increased costs for every new route $S^{\prime}_{V_{j}}[r_{i}]$. To illustrate, assume $A_{j}$ for vehicle $V_{j}$ has only two requests $r_{x}$ and $r_{y}$, its location is $loc(V_{j})$ and its current route has $r_{x}$ already inserted as: $[loc(V_{j}), o_{x}, d_{x}]$. Then, $V_{j}$ picks $S^{\prime}_{V_{j}}$ of inserting $r_{y}$ into its current route, such that:
${\small
 \text{cost}(V_{j}, S^{\prime}_{V_{j}}[r_{y}] ) = \text{min} [ w(loc(V_{j}), o_{x}, o_{y}, d_{x}, d_{y}), }\\ {\small 
 w(loc(V_{j}), o_{y}, o_{x}, d_{x}, d_{y}),
w(loc(V_{j}), o_{y}, o_{x}, d_{y}, d_{x}),}\\ {\small 
w(loc(V_{j}), o_{x}, o_{y}, d_{y}, d_{x}),
w(loc(V_{j}), o_{x}, d_{x}, o_{y}, d_{y}), }\\ {\small 
w(loc(V_{j}), o_{y}, d_{y}, o_{x}, d_{x}) ]}
$.
Note that the last two optional routes complete one request before serving the other, hence they do not fit into the \textit{ride-sharing} category. However, we still consider them as we optimize for the fleet's overall profits and total travel distance. Also, note that these two routes will still serve both requests and thus would not affect the overall acceptance rate of our algorithm. They may just increase the customers' waiting time a little, however, we show in our results that the customers' waiting time is very reasonable. Note that if this was the first allocation made to this vehicle, then the first request will be just added to its currently empty route. Otherwise, it will be dealt with (like all requests in the list) by following the insertion operation above. Then, the cost of serving all $k$ requests in matching $A_{j}$ is in turn defined as: $
 \text{cost}(V_{j}, A_{j}) = \sum_{r_{i} \in M_{j}} \text{cost}(V_{j}, S^{\prime}_{V_{j}}[r_{i}])$.
Finally, this phase works in a distributed fashion where each vehicle minimizes its travel cost, following Algorithm \ref{insert}. This distributed procedure goes on a customer-by-customer basis as follows:
\begin{itemize}[leftmargin=*]
\vspace{-0.05in}
\item Each vehicle $V_{j}$ receives its initial matchings list $A_{j}$, and an initial price $P_{init}$ (as explained in Section \ref{init_price}) associated with each request $r_{i}$ in that list. This initial matchings list is sorted ascendingly based on proximity to vehicle $V_{j}$.
\item Then, each vehicle $V_{j}$ considers each $r_{i} \in A_{j}$, in order of proximity, inserting into its current route. For each request $r_{i}$, the vehicle arrives at the minimum $\text{cost}(V_{j}, S^{\prime}_{V_{j}}[r_{y}] )$ of insertion into its current route (as described above). 
\item Now, given the initial price associated with this request $P_{init}$ and the the new route $S^{\prime}_{V_{j}}[r_{y}]$ (which may involve detours to serve this request), the vehicle can re-calculate the pricing to account for any extra distance, by feeding the new distance into Eq. \eqref{init} in Section \ref{init_price}.
\item Afterwards, drivers will then modify the pricing based on the Q-values of the driver's dispatch-to locations. Using the DQN dispatch policy on a regular basis, drivers have gained insight about which destinations can yield him/her a higher profit.  So, they weigh in on their utility function and propose a new pricing to the customer (explained in Section \ref{propose}). 
\item Finally, the customer(s) can accept or reject based on his/her utility function as explained in Section \ref{decide}. If a customer accepts, the vehicle updates its route $S_{V_{j}}$ to be $S^{\prime}_{V_{j}}[r_{y}]$, otherwise $S_{V_{j}}$ remains unchanged. The vehicle then proceeds to the next customer and repeats the process. Rejected requests will be fed back into the system to be considered in the matching process initiated in the next timestep for other/same vehicles.
\end{itemize}

The key idea in the proposed matching algorithm is that the pricing of the passenger depends on the distance in the route based on the previous matched passengers. Thus, if the new passenger is going in the opposite direction, the distance will be larger leading to higher price, and we expect that the passenger would likely not accept the high price as compared to another vehicle that is going towards that direction. Of course, if the customer is willing to pay the high price, he/she will be matched. Thus, the increased pricing prioritizes passengers to be matched if their routes have intersections as opposed to if they are going in opposing directions. 

\begin{algorithm}
	\caption{Insertion-based Route Planning} \label {insert}
		\fontsmall
		\begin{algorithmic}[1]
		\STATE \textbf{Input: } Vehicle $V_{j}$, its current route $S_{V_{j}}$, a request $r_{i} = (o_{i}, d_{i})$ and weighted graph G with pre-calculated trajectories using OSRM model.
		\STATE \textbf{Output: } Route $S^{\prime}_{V_{j}}$ after insertion, with minimum cost($V_{j}, S^{\prime}_{V_{j}}$). 
		\IF { $S_{V_{j}}$ is \textit{empty}}
			\STATE $S^{\prime}_{V_{j}}  \gets [loc(V_{j}), o_{i}, d_{i}]$.
			\STATE cost($V_{j}, S^{\prime}_{V_{j}}$) = $w(S^{\prime}_{V_{j}}) $.
			\STATE \textbf{Return} $S^{\prime}_{V_{j}}$, cost($V_{j}, S^{\prime}_{V_{j}}$)
		\ENDIF
		\STATE \textbf{Initialize} $S^{\prime \prime}_{V_{j}} = S_{V_{j}}$, $Pos[o_{i}] = \text{NULL}$, $\text{cost}_{min} = +\infty$.
		\FOR {each $x$ in $1$ to $|S_{V_{j}}|$}
			\STATE $S^{x}_{V_{j}}\; :=$ \textbf{Insert} $o_{i}$ at $x-th$ in $S_{V_{j}}$.
			\STATE \textbf{Calculate} cost($V_{j}, S^{x}_{V_{j}}$) = $w(S^{x}_{V_{j}})$.
			\IF {cost($V_{j}, S^{x}_{V_{j}}) < \text{cost}_{min}$}
				\STATE $\text{cost}_{min} \gets$ cost($V_{j}, S^{x}_{V_{j}}$).
				\STATE $Pos[o_{i}] \gets x$, $S^{\prime \prime}_{V_{j}} \gets S^{x}_{V_{j}}$.
			\ENDIF
		\ENDFOR
		\STATE $S^{\prime}_{V_{j}} = S^{\prime \prime}_{V_{j}}$, $\text{cost}_{min} = +\infty$.
		\FOR {each $y$ in $Pos[o_{i}] + 1$ to $|S^{\prime \prime}_{V_{j}}|$}
			\STATE $S^{y}_{V_{j}}\; :=$ \textbf{Insert} $d_{i}$ at $y-th$ in $S^{\prime \prime}_{V_{j}}$.
			\STATE \textbf{Calculate} cost($V_{j}, S^{y}_{V_{j}}$) = $w(S^{y}_{V_{j}})$.
			\IF {cost($V_{j}, S^{y}_{V_{j}}) < \text{cost}_{min}$}
				\STATE $\text{cost}_{min} \gets$ cost($V_{j}, S^{y}_{V_{j}}$).
				\STATE $S^{\prime}_{V_{j}} \gets S^{y}_{V_{j}}$, cost($V_{j}, S^{\prime}_{V_{j}}) \gets \text{cost}_{min} $.
			\ENDIF
		\ENDFOR
		\STATE \textbf{Return} $S^{\prime}_{V_{j}}$, cost($V_{j}, S^{\prime}_{V_{j}}$)
		\end{algorithmic} 
\end{algorithm}

\vspace{-0.1in}
\subsubsection*{\textbf{Complexity Analysis}} 
The complexity of the insertion operation is discussed in Appendix \ref{analysis}. Here we discuss checking the route feasibility in $O(1)$. For a route to be feasible, for each request $r_{i}$ in this route, $o_{i}$ has to come before $d_{i}$. Therefore, to further reduce the computation needed, we first find the optimal position $Pos[o_{i}]$ to insert $o_{i}$ and then, find the optimal position $Pos[d_{i}]$ to insert $d_{i}$ we only consider positions starting from $Pos[o_{i}] + 1$. Therefore, we never have to check all permutations of positions, we only check $n^{2}$ options in the ride-sharing environment as we check the route feasibility in $O(1)$ time. This is further reflected in the capacity constraint defined in phase 1, where we borrow the idea of defining the smallest position to insert origin without violating the capacity constraint from \cite{6} as $Ps[l]$, the capacity constraint of vehicle $V_{j}$ will be satisfied if and only if: $Ps[l] \leq Pos[o_{i}]$.
Here, we need to guarantee that there does not exist any position $l \in (Pos[o_{i}], Pos[d_{i}])$ such that: $V^{j}_{C}[l] \; \geq \; \; (\mid r_{i-1} \mid - \mid r_{i} \mid)$, other than $Ps[l]$ to satisfy $Ps[l] \leq Pos[o_{i}]$ and thus abide by the capacity constraint.

\vspace{-0.02in}
\section{Distributed Pricing-based Ride-sharing (DPRS)} \label{pricing}
In this section, we explain our distributed pricing strategy that is built on top of our distributed ride-sharing environment. In our simulator setup, we consider various vehicle types with varying capacity, mileage,  price per mile-distance, price per waiting-minute, and base price for driver per trip denoted $B_{j}$. $B_{j}$ serves as the local minimum earning for the driver per trip.

\vspace{-0.1in}
\subsection{Initial Pricing} \label{init_price}
Initially, each vehicle calculates a price for each of its requests, taking into consideration several factors:
\begin{itemize}[leftmargin=*]
	\item The total trip distance, i.e., the distance till pickup plus the distance from pickup to drop off. Note that, this distance is composed of the weights of the $n$ edges that constitute the vehicle's optimal route from its current location to origin $o_{i}$ and then to destination $d_{i}$. This route is obtained through the insertion operation in Algorithm \ref{insert},  as the route that minimizes the DARM cost function after inserting request $r_{i}$  
	and subtracting the cost of the original route before insertion. Thus, in the optimization step,  the new cost($V_{j}, S^{\prime}_{V_{j}[r_{i}]}$) is plugged into the equation to get the updated pricing.
	\item Number of customers who share travelling a trip distance (whether all or part of it, which can be determined from the vehicle's path). For simplicity, we denote it by the vehicle $j$'s capacity $V^{j}_{C}[o_{i}]$ when it reaches the origin $o_{i}$ location of this request $r_{i}$ subtracted from its capacity when it reaches the destination $d_{i}$, $V^{j}_{C}[d_{i}]$. Thus, we define $V^{j}_{C}[r_{i}] = \; \mid V^{j}_{C}[d_{i}] - V^{j}_{C}[o_{i}] \mid$. 
	\item The cost for fuel consumption associated with this trip, denoted by [distance travelled$*(P_{gas}/M_{j})$,] where $P_{gas}$ represents the average gas price, and $M^{j}_{V}$ denotes the mileage for vehicle $j$ assigned to trip $i$.
	\item The waiting time experienced by the customer (or customers) associated with trip $i$ till pickup, denoted $T_{i}$.
\end{itemize}
The overall pricing equation for request $r_{i}$ is represented as:
\begin{equation}\label{init}
{\small
\begin{multlined}
 P_{init}[r_{i}] = B_{j} + \left[ \omega^{1}*\frac{cost(V_{j}, S_{V_{j}[r_{i}]})}{V^{j}_{C}[r_{i}]} \right] \; + \\ \; \left[ \omega^{2}* \left( \frac{cost(V_{j}, S_{V_{j}[r_{i}]})}{V^{j}_{C}[r_{i}]}*(\frac{P_{gas}}{M^{j}_{V}}) \right) \right] \; - \; \left[ \omega^{3}*T_{i} \right]
\end{multlined}}
\end{equation}
where 
$\omega^{1}$ is the price per mile distance according to the vehicle type. $\omega^{2}$ is set to 1 as it doesn't change across vehicles, what changes is the mileage in this factor. Finally, $\omega^{3}$ is the price per waiting minute that is influenced by the vehicle type, it is negative here as we want to minimize the waiting time for the customer. 
Our proposed algorithm will first use the initial price  according to cost($V_{j}, S_{V_{j}[r_{i}]}$) and notify the vehicle (or driver), who will then modify the pricing based on the updated route cost($V_{j}, S^{\prime}_{V_{j}[r_{i}]}$) as well as the Q-values of the driver's dispatch-to location (explained in Section \ref{propose}).

\vspace{-0.1in}
\subsection{Vehicles' Proposed Pricing} \label{propose}
The core intuition behind assessing the cost/benefit of picking up a passenger is in having knowledge over the supply-demand distribution over the city.  This is learnt by each driver through the dispatch policy that they follow once he/she enters the market. This dispatch policy aims to provide him/her with the best next dispatch action to make, which is predicted after weighing the expected discounted rewards (Q-values) associated with each possible move on the map using DQN (described in Section \ref{dqn_algo}). As a result of running such a policy every dispatch cycle (set to 5 minutes in our simulation), the driver gains the necessary insight about how the supply-demand is distributed over the city, and thus can make informed decisions on the pricing strategy that can yield him a higher profit. This decision-making process is captured as follows: With the knowledge of the expected discounted sum of rewards (Q-values) across the map, the vehicle:
\begin{itemize}[leftmargin=*]
	\item Ranks destinations on the map, in a descending manner, according to the expected discounted sum of rewards, which is obtained using the DQN's Q values.  This rank is denoted $\alpha$.
	\item Dynamically maintains a list of highest ranked $\lambda$ regions on the map, denoted as $L$. This list represents the potential hotspots on the map that are anticipated to maximize the driver's profits,  and thus are \textit{desired zones}.  It is continuously updated whenever the vehicle runs its dispatch policy.
	\item After the route planning optimization step, the driver re-calculates the initial pricing $P_{init}(r_{i})$ using the updated route $S^{\prime}_{V_{j}}$ , by plugging cost($V_{j}, S^{\prime}_{V_{j}[r_{i}]}$) (after subtracting the cost of the original route before insertion) into Eq. \eqref{init}. This is done to account for any detours required to serve this new request $r_{i}$, and thus enforces requests going in the same direction to be matched together.
	\item Knowing the request location $loc(r_{i}$), if $ \in L$, the driver uses the initial pricing suggested for this trip, denoted $P_{init}(r_{i})$.
	\item Otherwise, it would indicate that driver might end up in the middle of nowhere (i.e., region with low demand), and thus receives no more requests or at least drives idle for a long distance. Instead of just rejecting the request,  the driver suggests a higher price to the customer to make up for the cost incurred due to mobilizing to a region of low demand. The price increase is influenced by both the rank of the destination as well as the driver's own base price per trip $B_{j}$ as in Eq. \eqref{car}.  
\begin{equation} \label{car}
{\small
P(r_{i}) = \begin{cases}
P_{init}(r_{i}) &\text{if $loc(r_{i}) \in L$}\\
P_{init}(r_{i}) +  \\ [P_{init}(r_{i}) *\frac{\alpha_{loc(r_{i})}}{2}*B_{j}] &\text{otherwise}
\end{cases}}
\end{equation}
\end{itemize} 

\vspace{-0.1in}
\subsection{Customers' Decision Function}\label{decide}
Upon the vehicle's proposed price it becomes the customer's turn to make his/her own decision according to his/her set of preferences. In our algorithm, we consider various preferences for each customer that constitutes their utility function:
\begin{itemize}[leftmargin=*]
	\item Tolerance in waiting time: whether the customer is in a hurry and how much delay can be tolerated: denoted as delay/waiting time of trip $i$: $T_{i}$.
	\item Preference in car-pooling: whether the customer is willing to share this ride or would rather take the ride alone even if it means a higher price. This is captured based on the current capacity of vehicle $j$, denoted by $V^{j}_{C}$.
	\item Preference of vehicle type for their trip: whether he/she is willing to pay more in exchange for a more luxurious vehicle. The type of vehicle $j$ is denoted by $V^{j}_{T}$.
\end{itemize} 
Based on the aforementioned factors, the customer's utility for request/trip $i$ is formulated as:
\begin{equation} \label{customer1}
{\small
U_{i} = \left[ \omega^{4} * \frac{1}{V^{j}_{C}} \right] + \left[ \omega^{5} * \frac{1}{T_{i}} \right] + \left[ \omega^{6} * V^{j}_{T} \right]
}
\end{equation}

where $\omega^{4}, \omega^{5}$, and $\omega^{6}$ are the weights associated with each of the factors affecting the customer's overall utility. To add more flexibility, we introduce a customer's compromise threshold $\delta_{i}$ to represent how much the customer $i$ is willing to compromise in the decision-making process. Finally, the decision of customer $i$ to accept or reject, denoted by $C^{i}_{d}$, after receiving the final price $P(r_{i})$ for the trip $i$ is as follows:
\begin{equation}\label{customer2}
{\small
C^{i}_{d} = \begin{cases}
1 &\text{if $ U_{i} > P(r_{i}) - \delta_{i}$}\\
0 &\text{otherwise}
\end{cases}}
\end{equation}


\section{Distributed DQN Dispatching Approach}\label{dqn_algo}
We utilize a distributed DQN dispatch policy to re-balance idle vehicles to areas of predicted high demand and profits over the city, where they can better serve the demand and maximize their profits.  We utilize a reinforcement learning framework, with which we can learn the probabilistic dependence between vehicle actions and the reward function thereby optimizing our objective function. Idle vehicles get dispatched when they first enter the market or when they experience large idle times throughout our simulation.

At every time step $t$, idle vehicles observe the state of the environment, $s_{t, n}$, and perform inference on their trained DQN to predict a future reward $r_{t}$ associated with each dispatch-to location on their action space $a_{t, n}$ over the map. Based on this information, the agent takes an action that directs the vehicle to different dispatch zone where the expected discounted future reward is maximized, i.e., $\sum^{\infty}_{j = t} \eta^{j-t} r_{j}(a_{t}, s_{t})$, where $\eta < 1$ is a time discount factor. This ultimately improves the fleet utilization. The overall flow of this framework is explained in Algo. \ref{dqn_brief}.  Lines $6-8$ of the algorithm specifically describe the best action that a given vehicle $V_{j}$ infers from the trained Q-network given the state $s_{t,n}$ and set of possible actions $a_{t,n}$.

In our algorithm, we define the reward $r_{t}$ as a weighted sum of different performance components that reflect the objectives of our DQN agent. The decision variables are i) Dispatching of an available vehicle in zone $m$, $V_{j} \in v_{t,m}$ to another zone at time slot $t$, ii) if a vehicle $V_{j}$ is not full, decide $\gamma_{j,t}$ its availability for serving new customers at time slot $t$. 
The reward will be learnt from the environment for individual vehicles and then leveraged to optimize their decisions.  Thus, the overall system objective is optimized at each vehicle in the distributed transportation network.  Below, we explain the state, action, and reward for our dispatch policy:\\
\textbf{State Space:} The state variables are utilized to reflect the environment status and thus influence the reward feedback to the agents' actions. 
We combine all the environment data explained in Section \ref{MP}: (1) $(X_{t}$: that keeps track of vehicles states: \textit{current zone of vehicle $v$, available seats, pick-up time, destination zone of each passenger.} (2)$V_{t:t+T}$: Supply prediction of the number of available vehicles at each zone for T time slots ahead, (3)$D_{t:t+T}$: Demand prediction at each zone for T time slots ahead. Thus, the state space at time $t$ is captured by three tuples combined in one vector as {\small $s_{t} = (X_{t}, V_{t:t+T}, D_{t:t+T})$}.  At each vehicle-request assignment, the simulator engine updates the state space tuple with the expected pick-up time, source, and destination data.  The three-tuple state variables $s_{t}$ are passed as an input to the DQN input layer which consequently outputs the best action to be taken.\\
\textbf{Action Space:} $a_{t}^{n}$ denotes the action taken by vehicle $n$ at time step $t$. In our simulator, the vehicle can move (vertically or horizontally) at most 7 cells,  it can move to any of the 14 vertical (7 up and 7 down) and 14 horizontal (7 left and 7 right) cells and hence the action space is limited to these cells. This results in a 15x15 action space $a_{t, n}$ for each vehicle as a vehicle can move to any of these cells or remain in its own cell. After the vehicle decides on which cell to go to using DQN, it uses the shortest optimal route to reach its next stop.  \\
\textbf{Reward:} 
The reward function (in Eq. \eqref{individual}) is a weighted sum of the following terms: (1) $C_{t,n}$: number of customers served by vehicle $n$ at time $t$, (2) dispatch time, $T_{t,n}^{D}$, taken by vehicle $n$ at time $t$ to go to zone $m$ or take detours to pick up extra requests. This term discourages the agent from picking up additional orders without considering the delay for on-board passengers. (3) $T_{t,n}^{E}$ denotes the sum of additional time vehicle $n$ takes at time $t$ to serve additional passengers, (4) profit for vehicle $n$ at time $t$: $\mathbb{P}_{t,n}$, and (5) $\max (e_{t,n} - e_{t-1,n}, 0)$ this term addresses the objective of minimizing the number of vehicles at time $t$ to improve vehicle utilization.
\vspace{-.1in}
\begin{equation} \label{individual}
\begin{multlined}
r_{t, n} = \beta_{1} C_{t,n} + \beta_{2} T_{t,n}^{D} + \beta_{3} T_{t,n}^{E} + \\ \beta_{4} \mathbb{P}_{t,n} + \beta_{5} [\text{max}(e_{t,n} - e_{t-1,n}, 0)]
\end{multlined}
\end{equation}
We define the derivation of each component of our reward function in Appendix \ref{dispatch}. 
Although we are minimizing the number of active vehicles in time step $t$, if the total distance or the total trip time of the passengers increase, it would be beneficial to use an unoccupied vehicle instead of having existing passengers encounter a large undesired delay.  
The details of learning these Q-values associated with the action space is provided in Appendix \ref{learn}, and the architecture of our Deep Q-Network is presented in Appendix \ref{DQN_Arch}.

While the primary role of the DQN is to act as a means of dispatching idle vehicles, it contains useful signals on future anticipated demand that is utilized by other components of our method including DARM matching and DPRS Pricing. Note that the profits term  added to the reward function makes the output expected discounted rewards (Q-values) associated with each possible move on the map, a good reflection of the expected earnings gained when heading to these locations. This gives drivers an insight about the supply-demand distribution over the city which is essential in making knowledgeable decisions when it comes to ranking their 
potential hotspots, and thus making the corresponding route planning and pricing decisions (Section \ref{darm} and Section \ref{propose}, respectively).
\begin{algorithm}
	\caption{Dispatching using DQN} \label {dqn_brief}
	\begin{minipage}{.9\linewidth}
		\fontsmall
		\begin{algorithmic}[1]
		\STATE \textbf{Input: } $X_{t}, V_{t:t+T}, D_{t:t+T}$.
		\STATE \textbf{Output: } Dispatch Decisions.
		\STATE \textbf{Fetch} all idle vehicles $\gets V_{\textit{Idle}}$.
		\FOR{each vehicle $V_{j} \in V_{\textit{Idle}}$ }
			\STATE \textbf{Construct} a state vector $s_{(t, n)} = (X_{t}, V_{t:t+T}, D_{t:t+T})$.
			\STATE \textbf{Push} state vector to the Deep Q-Network.
			\STATE \textbf{Get} the best dispatch action $a_{(t, j)} = \textit{argmax}[Q(s_{(t, n)}, a, \theta )]$. 
			\STATE \textbf{Get} the destination zone $Z_{(t,j)}$ based on action $a_{(t, j)}$.
			\STATE \textbf{Update} dispatch decisions by adding $(j, Z_{(t,j)})$.
		\ENDFOR
		\STATE \textbf{Return} Dispatch Locations $\forall$  $V_{j} \in V_{\textit{Idle}}$
		\end{algorithmic} 
	\end{minipage}
\end{algorithm}

\vspace{-0.1in}
\section{Experimental Results} \label{results}
\subsection{Simulator Setup}
In our simulator, we used the road network of the New York Metropolitan area along with a real public dataset of taxi trips in NY \cite{10}.  For each trip, we obtain the pick-up time, passenger count, origin location, and drop-off location. We use this trip information to construct travel requests demand prediction model as well. We start by populating vehicles over the city, randomly assigning each vehicle a type and an initial location. According to the type assigned to each vehicle, we set the accompanying features accordingly such as: maximum capacity, mileage, and price rates (per mile of travel distance $\omega^{1}$, and per waiting minute $\omega^{3}$). We initialize the number of vehicles, to $8000$. Note that, not all vehicles are populated at once, they are deployed incrementally into the market by each time step $t$.  We also defined a reject radius threshold for a customer request. Specifically, if there is no vehicle within a radius of 5km to serve a request, it is rejected. This simulator hosts each deep reinforcement learning agent which acts as a ridesharing vehicle that aims to maximize its reward: Eq. \eqref{individual}.
  
\vspace{-0.1in}
\subsection{DQN Training and Testing}
The fleet of autonomous vehicles was trained in a virtual environment that simulates urban traffic.  We consider the data of June 2016 for training and one week from July 2016 for evaluations. For each experiment, we trained our DQN neural networks using the data from the month of June 2016 for $20k$ epochs, which corresponds to a total of 14 days, and used the most recent 5000 experiences as a replay memory.  In Appendix \ref{learn}, Fig. \ref{fig:Learning Curve} shows the convergence of average Q-max during training.  Upon saving Q-network weights, we retrieve the weights to run testing on an additional 8 days from the month of July which corresponds to $10k$ epochs. Thus, T = 8 $\times$ 24 $\times$ 60 steps, where $\Delta t$ = 1 minute.  Also, we use \textit{Python} and \textit{Tensorflow} to implement our framework. Each vehicle has a maximum working time of $21$ hours per day, after which it exits the market.
To initialize the environment, we run the simulation for 20 minutes without dispatching the vehicles. Finally, we set $\beta_{1} = 10, \beta_{2} = 1, \beta_{3} = 5$, $\beta_{4} = 12$, $\beta_{5} = 8$, $\lambda = 10\% $, $\omega^{4} = 15, \omega^{5} = 1$, and $\omega^{6} = 4$. 

\begin{figure*}
\includegraphics[trim=0 0 70 30, width=0.93\textwidth]{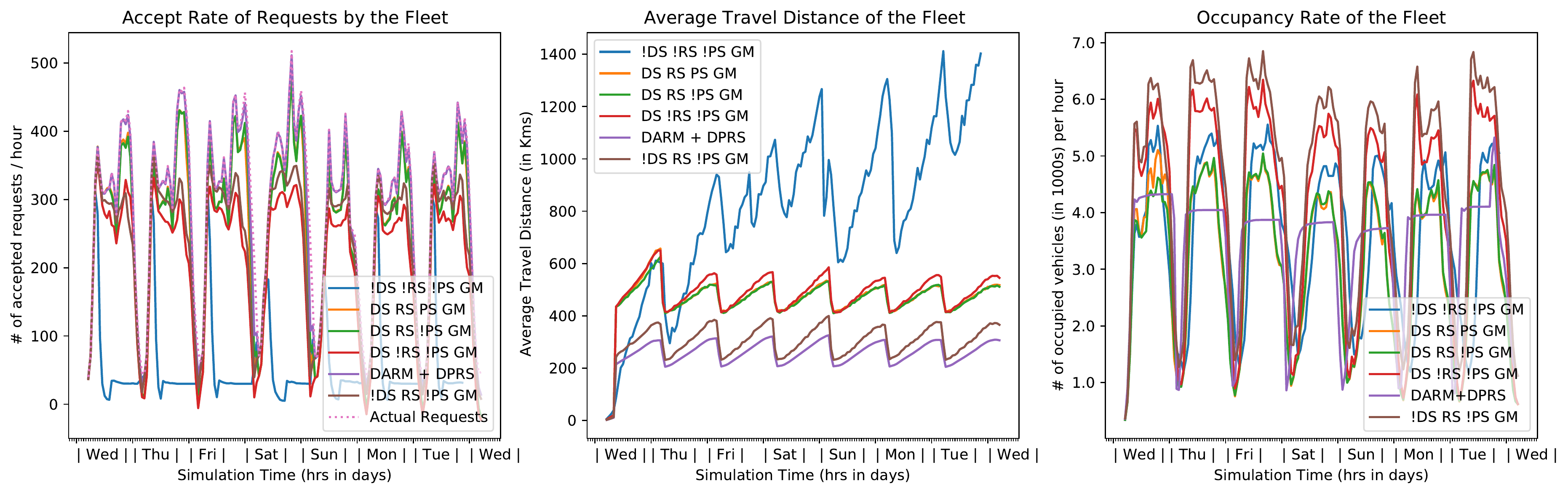} 
 \vspace{-0.1in}
\caption{Performance Metrics of the proposed algorithm and the  baselines \protect \footnotemark}\label{sumary} 
 \vspace{-0.1in}
\end{figure*}

\vspace{-0.12in}
\subsection{Performance Metrics}
\begin{figure*}
  \centering
 \begin{tabular}{cc}
 \begin{subfigure}{0.5\textwidth}
\includegraphics[trim=30 10 0 5, width=\textwidth]{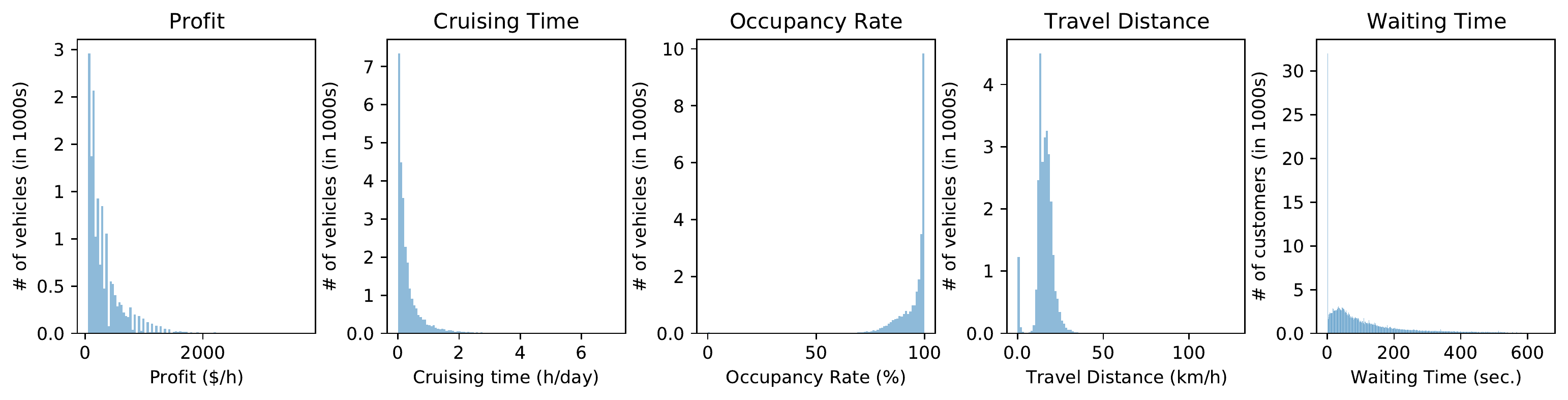} 
\caption{\small Performance Metrics of our Joint Framework \\ ``DARM + DPRS" }\label{darm_sumary}
\end{subfigure} \hfill
\begin{subfigure}{0.5\textwidth}
\includegraphics[trim=10 10 30 5, width=\textwidth]{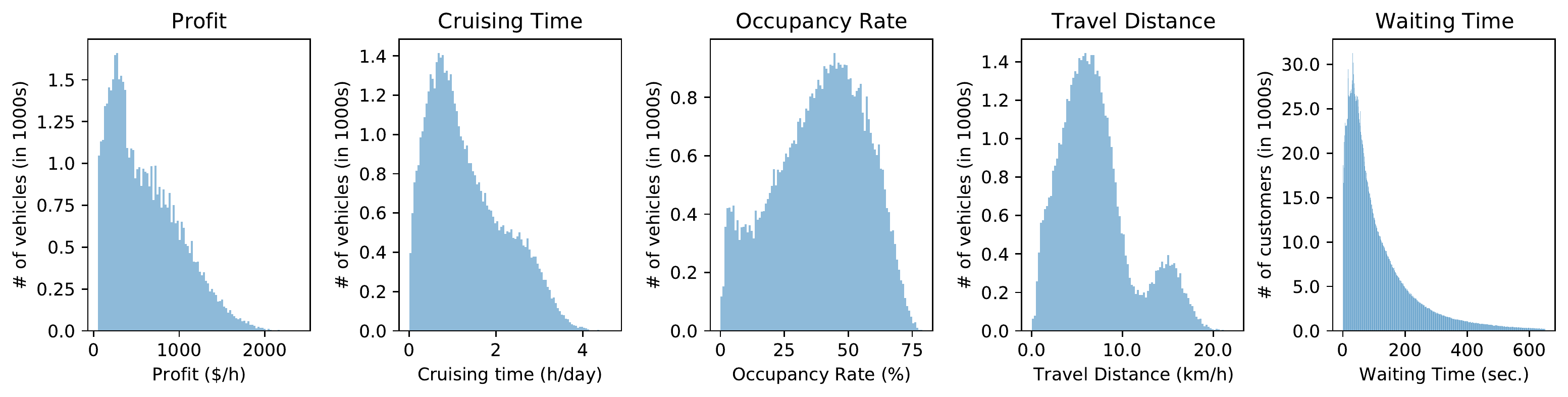} 
\caption{\small Performance Metrics of (D, RS, PS, GM) Baseline \\ ``DPRS with Greedy Matching"}\label{dprs_sumary}
\end{subfigure}  \\ [1em]
\begin{subfigure}{0.5\textwidth}
\includegraphics[trim=30 10 0 5, width=\textwidth]{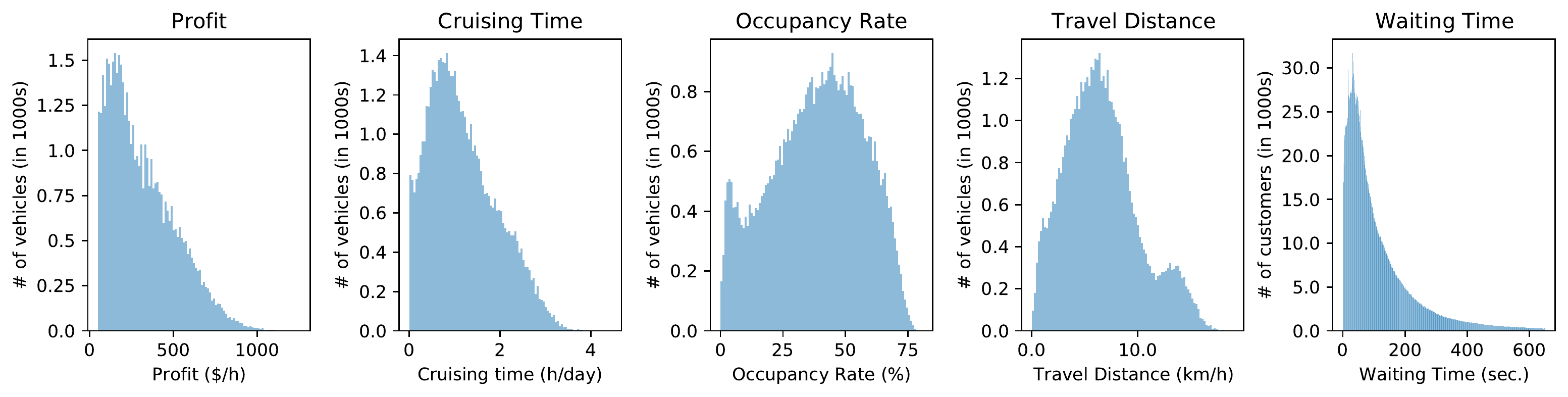} 
\caption{\small Performance Metrics of (D, RS, !PS, GM) Baseline \\ 	``Deep\_Pool in \cite{deep_pool}"}\label{pool_sumary}
\end{subfigure} \hfill
\begin{subfigure}{0.5\textwidth}
\includegraphics[trim=10 10 30 5, width=\textwidth]{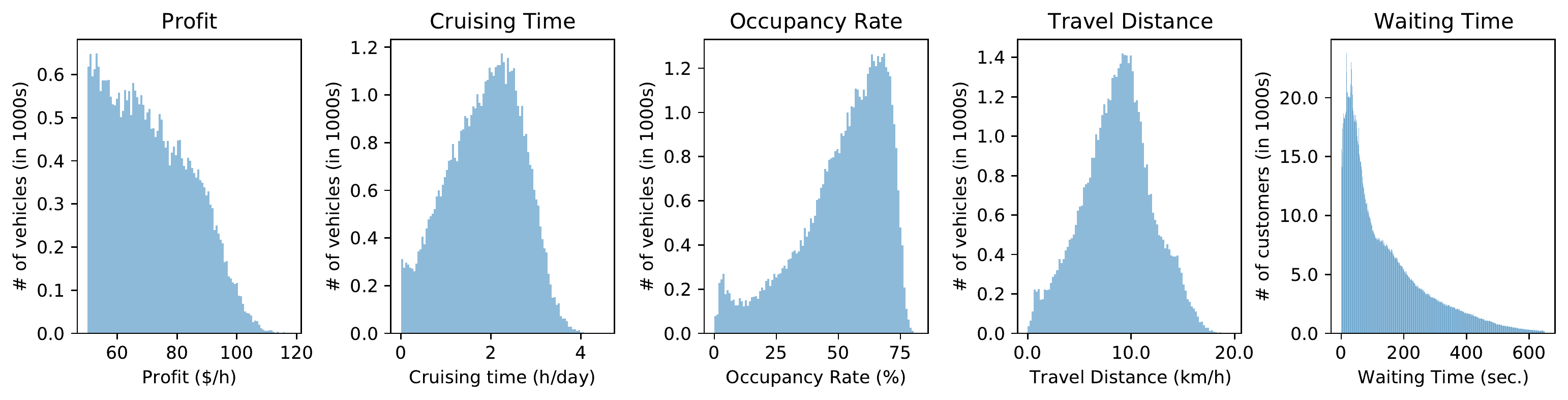} 
\caption{\small  Performance Metrics of (D, !RS, !PS, GM) Baseline \\ ``Dynamic Taxi Fleet Management in \cite{fleet_oda}" }\label{oda_sumary}
\end{subfigure}
\end{tabular}
 \vspace{-0.05in}
\caption{Histograms of Performance Metrics for the Proposed Algorithm and the Baselines \protect\footnotemark}
 \label{tab1}
 \vspace{-0.1in}
\end{figure*}

We breakdown the reward and utility functions, and investigate the performance for various baselines. Recall that we want to minimize the components of our reward in Eq. \eqref{individual}.  
\begin{itemize}[leftmargin = *]
\item \textbf{Accept rate: } we note that the supply-demand mismatch is reflected in our simulation through this metric.  Accept rate is defined as the ratio of successful pick-ups by the fleet to the total number of requests made in a given time slot. A high accept rate is a characteristic of a reliable mode of transportation.  With a high acceptance rate, our fleet is able to fulfill the passengers' transportation demands.
\item \textbf{Cruising (idle) time:} this metric represents the time at which a vehicle is neither occupied nor gaining profit but still incurring gasoline cost. 
\item \textbf{Occupancy Rate: } this metric captures the utilization rate of the fleet of vehicles, it keeps track of how many vehicles are deployed from the fleet to serve the demand. By minimizing the number of occupied vehicles, we achieve better utilization of individual vehicles in serving the demand. Given that (i) all baselines are catering to a similar volume of pickup orders, and (ii) if all baselines are achieving a similar accept rate,  a lower occupancy rate indicates that a fleet is able to minimize the number of vehicles on the street to serve the requests. Note that, in Fig.  \ref{tab1}, we also show the utilization of each individual vehicle (i.e., percentage of time the vehicle is occupied while in duty).
\item \textbf{Waiting Time: } this metric captures the time customers had to wait till they get picked up by a vehicle (shown in seconds). We note that wait time is an important metric for customer convenience with mobility-on-demand services.
\item \textbf{Profit: } this metric represents the net profit of each driver per hour of service, where the cost incurred by fuel consumption is subtracted from the revenue.  Note that, the Q-values depend on the pricing since the decisions made by customers and drivers impact the reward function through the profit term.  With high net profits, our framework is able to find a common ground that is profitable to both parties.
\item \textbf{Travel Distance: } this metric shows the number of kilometers traveled by each vehicle per hour of service, which gives a good reflection of the cost incurred by vehicles due to serving multiple ride requests.
\vspace{-0.12in}
\end{itemize}

\subsection{Baselines}
We compare our proposed framework (with dispatching, ride-sharing, our novel DPRS pricing strategy, and DARM approach for matching and route planning) against the following baselines to emphasize the distinct impact of each component: 
\begin{itemize}[leftmargin=*]
	\item No Dispatch, No Ride-sharing, No Pricing Strategy, Greedy Matching  (!D, !RS, !PS, GM): In this setting, vehicles don't get dispatched to areas with anticipated high demand, no matter how long they stay idle. Ride-sharing (pooling) is not allowed, every vehicle serves only one request at a time. Also, initial pricing is used and is accepted by both drivers and customers by default. For matching,  only the greedy initial matching is applied, no optimization takes place.
	\item No Dispatch with Ride-sharing but No Pricing Strategy, and with Greedy Matching (!D, RS, !PS, GM): similar to (!D, !RS, !PS) except that ride-sharing (pooling) is allowed, where vehicles can serve more than one request altogether.
	\item Dispatch with No Ride-sharing and No Pricing Strategy with Greedy Matching (D, !RS, !PS, GM): Here, vehicles are dispatched when idle but, ridesharing isn't allowed as \cite{fleet_oda}. 
	\item Dispatch with  Ridesharing but No Pricing Strategy, and with Greedy Matching (D, RS, !PS, GM): similar to (D, !RS, !PS, GM), but with ride-sharing allowed, as in DeepPool \cite{deep_pool}.
	\item Dispatch with  Ridesharing and Pricing Strategy, but with Greedy Matching (D, RS, PS, GM): similar to (D, RS, !PS, GM), but with applying our DPRS pricing strategy where customers and drivers are involved in the decision-making process. However,  only greedy matching is adopted here. 
\end{itemize}
\footnotetext{Enlarged Figure is provided in Fig. \ref{big} in Appendix D}
The proposed baselines aim to evaluate the effectiveness of each component of our framework.  Our proposed joint method incorporates both insertion-based route-planning and pricing strategy to involve both customers and drivers in the decision-making. As compared to the above baselines that don't adopt DARM, we hypothesize that our (DARM+DPRS) would be a more effective approach. Given that the core intuition of DARM is to group together rides that share route intersections to their destinations as opposed to rides heading to opposite-direction-destinations, we expect improvements in the number of rides served, profits,  travel distance, and occupancy rate. \\
Moreover, we have included baselines that do not consider DPRS, where Pricing Strategy is not adopted (!PS) to observe the effectiveness of our pricing framework. In these scenarios, involving drivers and customers in the decision-making process has the potential to burden the system causing an increase in the rejection rate and the number of vehicles utilized to serve the demand. However, we hypothesize that our DPRS together with our dispatch policy will be able to establish the balance and reach solutions that are profitable to both passengers and drivers. To validate our hypothesis, we investigate both the net profits of drivers as well as the passengers' waiting times. In addition, as compared to DPRS only baseline (D, RS, PS, GM), our joint framework (DARM + DPRS) is expected to further improve the fleet utilization, idle time and overall travel distance. 
In addition, we also include baselines that don't adopt dispatching policy where vehicles only mobilize according to the pickup locations of their requests as opposed to learning the supply-demand distribution of the city and mobilizing accordingly when they experience idle time. Comparing against this baseline shows the impact of our dispatch policy on improving profits, fleet utilization, accept rate, travel distance, and waiting times. 

Finally, since \cite{deep_pool} showed that (D, RS, PS, GM) performs better than the dispatch with minimium distance (DS-mRS) approach \cite{6} and the Centralized Receding Horizon Control (cRHC) approach \cite{miao2016taxi} in terms of idle times,  waiting times and fleet utilization, we do not consider these baselines. In (DS-mRS), ride-sharing is allowed where two riders are assigned to one vehicle so that the total driving distance is minimized.  However, in cRHC, the dispatch actions are taken to maximize the expected reward in a centralized manner as opposed to our distributed approach. 
	
We include baselines that don't adopt ridesharing to benchmark how much of the improvement in the aforementioned performance metrics is attributed to ride pooling instead of serving one ride at a time,  as opposed to the improvement due to the deployment of the other components of our framework (Dispatching Policy, DPRS and DARM). Note that DARM works in tandem with DPRS, as the pricing decisions impact the route planning procedure (i.e. pricing-based matching) and thus the improvement of profits, waiting times, idle times as well as travel distance is a result of deploying both approaches.

\vspace{-0.1in}
\subsection{Results Discussion}
We validate the necessity of each component of our framework using computational results.  From our simulation, we observe that the hypothesis for each baseline comparison has been supported for the most part by our experimental results.
In Fig. \ref{sumary}, we investigate the overall performance of our proposed framework in comparison to all other baselines.  We show the actual number of requests as the dotted pink line.  We observe that our joint DARM + DPRS framework ranks highest in the acceptance rate per hour (almost coincide with the actual requests curve) of $> 96\%$, followed by all the ridesharing-based baselines (at around $80\%$), while the non-ridesharing and the non-dispatching baselines come at the bottom of the list (with $< 55\%$). Clearly, our joint (DARM+DPRS) approach boosts the acceptance rate with an increase of $>15\%$ when compared to DPRS only [(D RS PS GM) baseline] and DeepPool [(D RS !PS GM) baseline] and $>40\%$ when compared to non-dispatching and/or non-ridesharing baselines. This proves our hypothesis that this improvement of $40\%$ is achieved by our joint (DARM + DPRS), while $15\%$ is fully attributed to our DARM approach (which makes use of DPRS in its optimization phase).\\
Contradictory to the expectation that involving drivers and customers in the decision-making process would increase the rejection rate and the number of vehicles utilized to serve the demand,  our joint (DARM+DPRS) has significantly low rejection rate. To further analyze the rejection rate due to DPRS,  we observe that the rejection rate made by customers (i.e., when a customer weighs in and rejects a ride) is fairly close to the naturally encountered rejection rate that occurs due to the unavailability of vehicles within the request's vicinity. 

\footnotetext{Enlarged figure is provided in Fig.  \ref{enlarged} in Appendix D.}

It is worth noting that the proposed method is able to deliver more requests while minimizing the number of occupied vehicles. Fig. \ref{sumary} shows a utilization/occupancy rate of around two-thirds of that of the rest of the baselines, saving one-third of the vehicles for serving new incoming requests which would -in turn- further increase the acceptance rate. Also, we had set the maximum number of vehicles in our simulator to $8000$, our approach utilizes only half of them ($4000$) to serve the demand with an acceptance rate $>96\%$ With the DPRS only [(D RS PS GM) baseline] and DeepPool [(D RS !PS GM) baseline], this increases to just below $5000$ vehicles. Again, an improvement of 1000 vehicles, only attributed to adopting our DARM approach. Without our dispatching policy, the number of occupied vehicles reaches more than 6000 that only covered $60\%$ of the demand. While, as expected, without ridesharing, more than $7500$ vehicles are utilized to serve only $60\%$ of the demand. Since our DARM + DPRS performs significantly better than DeepPool in both metrics; this makes DARM + DPRS superior to DeepPool, DS-mRS, and cRHC baselines.

We can further support our hypothesis by looking at the average travel distance of the fleet, we can see it is exploding (around $1300km$) with the non-dispatching non-ridesharing baseline, where the vehicle only serves one request at a time and is at the risk of encountering large cruising time while looking for a ride, without taking the right dispatch action to a zone where new requests can be found.  Just by adding ridesharing, travel distance dramatically decreases to around $500km$ [(!DS RS !PS GM) baseline] saving $>60\%$,  but in this case, the accept rate is only $55\%$. However, when adopting our dispatch policy [(DS RS !PS GM) baseline], travel distance reaches $700km$ saving $>50\%$, while serving $80\%$ of the demand.  Finally, with joint (DARM+DPRS), the overall travel distance falls at $300km$, saving $>80\%$ while serving $\approx 96\%$ of the demand. Therefore, a $30\%$ decrease in the overall travel distance is attributed to our DARM+DPRS framework.

Clearly, non-dispatching and non-ridesharing algorithms are shown to have poor utilization of resources, as they use a higher number of vehicles to serve the less amount of demand.  Therefore,  we exclude them due to their poor performance in Fig. \ref{sumary}., and we take a closer look at the ride-sharing based baselines that adopt a dispatch policy.  
Figure \ref{tab1} shows that the average profits for the drivers have significantly increased over time as compared to the other three protocols.  Thus, quantifying the individual drivers' preferred zones based on the learnt reward using DQN, could guarantee them a significant improvement in earnings that, in turn, helped make up for any extra encountered cost. This implies both the drivers and customers are achieving a compromise that is profitable and convenient to them. Specifically, the profits for our DARM + DPRS framework is almost double that of DPRS only [(D RS PS GM) baseline], and $3 - 4$ times that of DeepPool [(D RS !PS GM) baseline]. Without ride-sharing, profits are 10 times less 
than the average profit with our joint framework.

We will further show that our framework not only enhances the overall fleet utilization, but also the utilization of each individual vehicle (i.e., percentage of time the vehicle is occupied).  Similarly, for the travel distance metric. Fig. \ref{darm_sumary} shows that our joint framework minimizes the cruising time, where vehicles are idle, and thus minimizes the extra travel distance as well as extra gasoline cost.  On average, vehicles' idle time is within a minimal range ($1 - 2$) hours for DARM + DPRS framework. Knowing that vehicles' working time is at most $21$ hours per day, we observe that $> 85 \%$ of the allowed vehicles per day, experience idle time $< 2$ hours, which is $< 10\%$ of their total working time. This metric is almost doubled for all other three baselines. This proves that most of the improvement in fleet utilization is due to our DARM framework.  This is also reflected in the occupancy rate metric, which is defined as the percentage of time where vehicles are occupied while on duty. Fig. \ref{darm_sumary} shows that $6k - 8k$ vehicles are between $80\% - 100\%$ occupied, and hence proves that our framework significantly improves the utilization of each individual vehicle as well as the whole fleet.

However,  Fig. \ref{tab1} shows that the average travel distance of DARM + DPRS is slightly higher, ranging between $10 - 30$ km per hour, compared to $5 - 25$ km for the other baselines. Most of the $4000$ vehicles deployed by (DARM+DPRS) fall within $10-20$ km which is very reasonable given the $15-40\%$ extra demand that they serve. Since DARM + DPRS  provides significantly higher profits for drivers,  it comes at the cost of a slight increase in travel distance, which is an advantage of DARM + DPRS. Note that, the two policies with the lowest travel distance in Fig \ref{pool_sumary}, and \ref{oda_sumary} are not involving vehicles or customers in the decision-making process which explains why vehicles have lower travel distances on average. However, we can observe that the non-ridesharing protocols are not efficient as they result in lower profit margins and higher customers' waiting time.  Moreover, we emphasize that non-dispatching protocols yield higher idle time for the vehicles as they might spend a large amount of time being idle and they never get dispatched to higher demand areas. In contrast, non-ride-sharing protocols yield lower idle time, but they are still inefficient as vehicles spend more time on duty while serving a lower number of customers than the ride-sharing protocols. 

Compared to both DeepPool [(D RS !PS GM) baseline] and DPRS only [(D RS PS GM) baseline], the waiting time per request is significantly lower for DARM + DPRS approach. As shown in Fig. \ref{tab1}, the waiting time for customers reduces overtime to $< 1 $ minute. On average, the response time (time till customer is picked up) of our framework is $< 200$ sec ($\approx$ 3 minutes), which is almost half that of the other three policies. Note that, the framework in Figure \ref{oda_sumary} is a non-ridesharing framework which should have had a lower response time as the vehicles directly head to the customer to be served without having to pickup other customers on the way. However, our framework shows a similar response time, with a majority of customers experiencing waiting time $\approx$ 1 minute.


\vspace{-0.1in}
\section{Conclusion} \label{conc}
In this paper, we detailed two novel approaches\textemdash Demand-Aware and Pricing-based Matching and route planning (DARM) framework and Distributed Pricing approach for Ride-Sharing with pooling (DPRS)\textemdash that generate ideal routes on-the-fly and involve both customers as well as drivers in the decision-making processes of ride acceptance/rejection and pricing. Agents' decision-making process is informed by utility functions that aim to achieve the maximum profit for both drivers and customers. The utility functions also account for the fuel costs, waiting time, and passenger's spending power  to compute the reward.  These novel DARM and DPRS methodologies are also integrated via a Deep Q-network (DQN) based dispatch algorithm where the profits influence the dispatch and the Q-values impact the pricing and thus matching. 
Contradictory to an expected high rejection rate when agents are given a choice to reject rides, experimental results show that the rejection rate is significantly low for (DARM + DPRS) framework. Given the maximum number of vehicles (8000 vehicles) populated in the simulation, our framework only uses 50\% of the vehicles to accept and serve the demand of up to 96\% of the requests. When compared with no ride-sharing baselines, our framework provides 10 times more profits. Experiments also show that vehicle idle time (cruising without passengers) is reduced to under two hours and 80\% - 100\% of the vehicles are occupied all the time. 
Our model-free DARM + DPRS framework can be extended to large-scale ride-sharing protocols due to distributed decision making for different vehicles reducing the decision space significantly. 

Extension of this work to consider travel-time uncertainty where riders can change their ride information on-the-fly \cite{li2020ride}, and to compute global plans for a flexible ridesharing with different objectives \cite{armant2020fast} is left as future work. Additional future directions of extending this work are: including capabilities of a joint delivery system for passengers and goods as in \cite{manchella2020flexpool} (considered in part in \cite{manchella2020passgoodpool}), or using multi-hop routing of passengers as in \cite{singh2019distributed} and transit services as in \cite{ma2019dynamic} for efficient fleet utilization. 


\vspace{-0.1in}
\bibliographystyle{IEEEtran}
\bibliography{IEEEabrv,IEEEexample}
\newpage
\clearpage
\appendices
\setcounter{page}{1}

\section{Control Unit Components:}\label{models}
\subsection{OSRM and ETA Model}
We construct a region graph relying on the New York city map, obtained from OpenStreetMap \cite{osm}. Also, we construct a directed graph as the road network by partitioning the city into small service area 212 x 219 bin locations of size 150m x 150m. We find the closest edge-nodes to the source and destination and then search for the shortest path between them. To estimate the minimal travel time for every pair of nodes, we need to find the travel time between every two nodes/locations on the graph. To learn that, we build a fully connected neural network using historical trip data as an input and the travel time as output. The fully connected multi-layer perception network consists of two hidden layers with width of 64 units and rectifier nonlinearity. The output of this neural network gives the expected time between zones. While this model is relatively simple (contains only two hidden layers), our goal is to achieve a reasonable accurate estimation of dispatch times, with short running time. Finally, if there is no vehicle in the range of $5 \; km^{2}$, the request is considered rejected.

\vspace{-0.1in}
\subsection{Demand Prediction Model}
\begin{figure}
	\vspace{-2.5em}
	\begin{center}
		\includegraphics[width=0.5\textwidth]{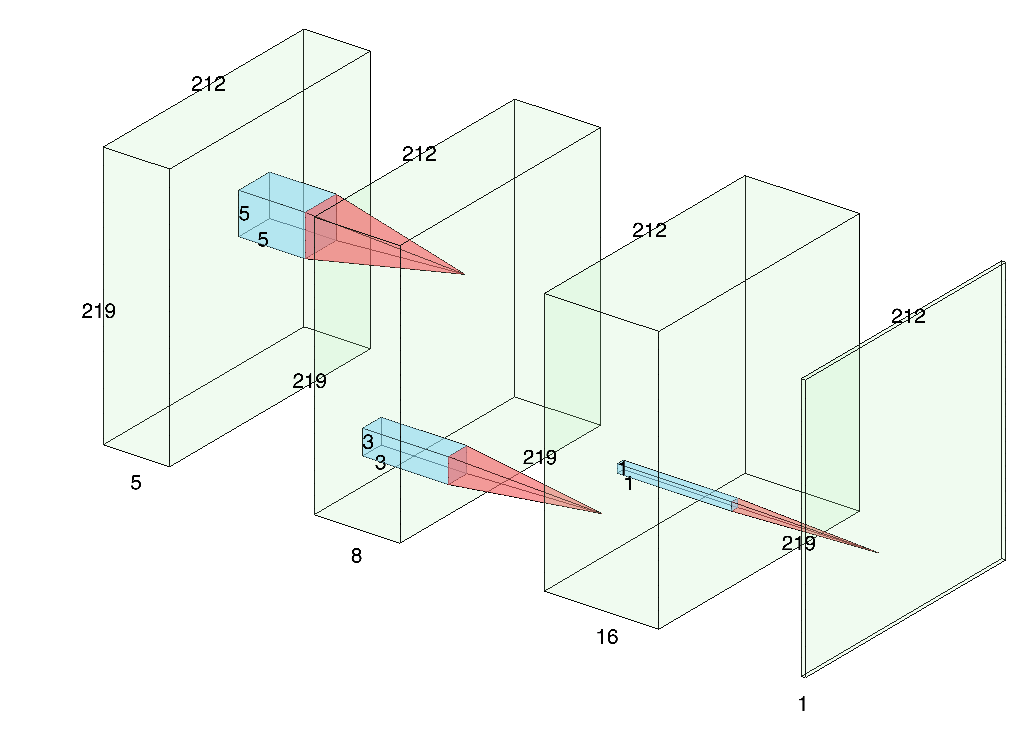}
		\vspace{-3em}
		\caption{The architecture of the Convolution Neural Net used for Demand Prediction.} \label{demand}
	\end{center}
\end{figure}

We use a convolutional neural network to predict future demand. Figure \ref{demand} shows the architecture of this Conv-Net. The output of the network is a $212 x 219$ image such that each pixel represents the expected number of ride requests in a given zone for $30$ minutes ahead. The network input consists of two planes of actual demand of the last two steps. The size of each plane is 212 x 219. The first hidden layer convolves $16$ filters of size 5 x 5 with a rectifier nonlinearity, while the second layer convolves 32 filters of 3 x 3 with a rectifier nonlinearity. The output layer convolves 1 filter of size 1 x 1 followed by a rectifier nonlinear function.

\section{Insertion Operation} \label{analysis}
\subsection{Complexity Analysis:} As the basic insertion algorithm needs to check every possible insertion position, which is $O(n^{2})$, where $n$ is the number of potential requests per vehicle. Then, for every insertion position pair, calculates the new cost of the new route which is $O(n)$, it results in complexity of $O(n^{3})$. Besides, to realize a constant-time cost calculation in DARM, we propose a list to pre-derive all the needed cases such that any situation can look up the list to find the increased cost in constant time $O(1)$. Based on the granularity of time values (e.g., at least 30 seconds as the unit time for drop-off), we can dynamically derive all the situations according to finite cases of time delay after insertion. Therefore, the routes (with their associated costs) pre-calculation step done using our OSRM engine, provides us with fast routing and constant-time computation $O(1)$, thus reduces the complexity of our algorithm from $O(n^{3})$ to $O(n^{2})$. This can be further improved by using dynamic programming and adopting the approach in \cite{xu2020efficient} to achieve $O(n)$.

\section{DQN Dispatching Algorithm} \label{dqn}
\subsection{DQN Architecture} \label{DQN_Arch}

\begin{figure}
	\vspace{-2.5em}
	\begin{center}
		\includegraphics[scale=0.3]{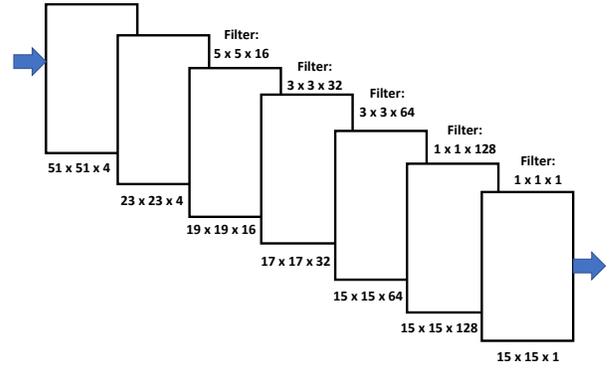}
		\vspace{-3em}
		\caption{The architecture of the Q-network. The output represents the Q-value for each possible movement/dispatch} \label{dqn_arch}
	\end{center}
\end{figure}

Figure \ref{dqn_arch} demonstrates the architecture of the Q-network. The output represents the Q-value for each possible movement/dispatch. In our simulator, the service area is divided into 43x44, cells each of size 800mx800m. The vehicle can move (vertically or horizontally) at most 7 cells, and hence the action space is limited to these cells. A vehicle can move to any of the 14 vertical (7 up and 7 down) and 14 horizontal (7 left and 7 right). This results in a 15x15 map for each vehicle as a vehicle can move to any of the 14 cells or it can remain in its own cell. The input to the neural network consists of the state representation, demand and supply, while the output is the Q-values for each possible action/move (15 moves). The input consists of a stack of four feature planes of demand and supply heat map images each of size 51x51. In particular, first plane includes the predicted number of ride requests next 30 minutes in each region, while the three other planes provide the expected number of available vehicles in each region in 0; 15 and 30 minutes. Before passing demand and supply images into the network, different sizes of average pooling with stride (1, 1) to the heat maps are applied, resulting in 23 x 23 x 4 feature maps. The first hidden layer convolves 16 filters of size 5x5 followed by a rectifier non-linearity activation. The second and third hidden layers convolve 32 and 64 filters of size 3x3 applied a rectifier non-linearity. Then, the output of the third layer is passed to another convolutional layer of size 15 x 15 x 128. The output layer is of size 15 x 15 x 1 and convolves one filter of size 1x1. Since reinforcement learning is unstable for nonlinear approximations such as the neural network, due to correlations between the action-value, we use experience replay to overcome this issue. Since every vehicle runs its own DQN policy, the environment during training changes over time from the perspective of individual vehicles.

\vspace{-0.1in}
\subsection{DQN Dispatch Agent} \label{dispatch}
Initially, we define the overall objectives of the dispatcher as well as the decision variables in Section \ref{dqn_algo}. Below, we present the how each of these objectives is calculated and represented with a corresponding term:

\begin{enumerate}[leftmargin=*]
	\item Minimize the supply-demand mismatch, recall that $v_{t,m}$, and $\bar{d}_{t,m}$ denotes the number of available vehicles, and the anticipated demand respectively at time step $t$ in zone $m$. We want to minimize their difference over all $M$ zones, therefore, we get:
	\vspace{-0.1in}
	\begin{equation} \label{total}
	\text{diff}_{t} = \sum^{M}_{m = 1} (\bar{d}_{t,m} - v_{t,m})
	\end{equation}
	The reward will be learnt from the environment for individual vehicles, therefore, we map this term for individual vehicles. When vehicle serves more requests, the difference between supply and demand is minimized, and helps satisfy the demand of the zone it is located in. Therefore, we can get the total number of customers served by vehicle $n$ at time step $t$:
	\vspace{-0.1in}
	\begin{equation}
	\begin{multlined}
	\text{C}_{t,n} = \sum^{M}_{m = 1} v^{n}_{t,m} \: \: \: \;	(\text{where }v_{t,m} = 1 \text{ when } v_{t,m} < \bar{d}_{t,m}) \\
	\text{where     } \sum^{M}_{m = 1} v^{n}_{t,m} = 1 \: \: \: \; (\gamma_{n,t,m} \in \{0,1\} \; \;  \text{where } \: n \in v_{t,m}) \; \; \; \; \; \; \; \; \; \;
	\end{multlined}
	\end{equation}
	
	\item Minimize the dispatch time, which refers to the expected travel time of vehicle $V_{j}$ to go zone $m$ at time step $t$, denoted by $h^{j}_{t,m}$. We calculate this time from the location of vehicle $V_{j}$ at time $t$ which is already included in the state variable $X_{t, j}$. Idle vehicles get dispatched to different zones (where anticipated demand is high) than their current zones (even if they do not have any new requests yet), in order to pick up new customers in the future. Since we want to minimize over all available vehicles $N$ over all zones $M$ within time $t$, we get the total dispatch time, $T^{D}_{t}$ as follows:
	\vspace{-0.1in}
	\begin{equation}
	T^{D}_{t} = \sum^{N}_{n = 1} \sum^{M}_{m = 1} h^{n}_{t,m} \; \; \; \; \{\forall \: n \in v_{t,m}\}
	\end{equation}
	For individual vehicles, considering the neighboring vehicles' locations while making their decision, we get for vehicle $n$ at time step $t$:
	\vspace{-0.1in}
	\begin{equation}
	T^{D}_{t, n} = \sum^{M}_{m = 1} h^{n}_{t,m} \; \; \; \; \{\text{where } \: n \in v_{t,m}\}
	\end{equation}
	\item Minimize the difference in times that the vehicle would have taken if it only serves one customer and the time it would take for car-pooling. For vehicles that participate in ride-sharing, an extra travel time may be incurred due to (1) either taking a detour to pickup an extra customer or (2) after picking up a new customer, the new optimal route based on all destinations might incur extra travel time to accommodate the new customers. This will also imply that customers already on-board will encounter extra delay. Therefore, that difference in time needs to be minimized, otherwise both customers and drivers would be disincentivized to car-pool. Let $t'$ be the total time elapsed after the passenger $l$ has requested the ride, $t_{n,l}$ be the travel time that vehicle $n$ would have been taken if it only served rider $l$, and $\tilde{t}_{n,l}$ be the updated time the vehicle $n$ will now take to drop off passenger $l$ because of the detour and/or picking up a new customer at time $t$. Note that $\tilde{t}_{n,l}$ is updated every time a new customer is added. Therefore, for vehicle $n$, rider $l$ at time step $t$, we want to minimize: $ \xi_{t,n,l} =  t' + \tilde{t}_{n,l} - t_{n,l}$. But for vehicle $n$, we want to minimize over all of its passengers, thus: $\sum^{\cup_{n}}_{l = 1} \xi_{t,n,l}$, where $\cup_{n}$ is the total number of chosen users for pooling at vehicle $n$ till time $t$. Note that $\cup_{n}$ is not known apriori, but will be adapted dynamically in the DQN policy. It will also vary as the passengers are picked or dropped by vehicle $n$. We want to optimize over all $N$ vehicles, therefore, the total extra travel time can be represented as:
	\vspace{-0.1in}
	\begin{equation}
	\Delta t = \sum^{N}_{n = 1} \sum^{\cup_{n}}_{l = 1} \xi_{t,n,l}.
	\end{equation}
	For individual vehicles, extra travel time for vehicle $n$ at time step $t$ becomes:
	\begin{equation}
	T^{E}_{t,n} = \sum^{\cup_{n}}_{l = 1} \xi_{t,n,l}.
	\end{equation}
	\item Maximize the fleet profits. This is calculated as the average earnings $E_{t}$ minus the average cost of all vehicles. Cost is calculated by dividing the total travel distance of vehicle $V_{j}$ by its mileage, and multiplied by the average gas price $P_{G}$. Therefore, the average profits for the whole fleet can be represented as:
	\vspace{-0.1in}
	\begin{equation}
	\mathbb{P}_{t} = \sum^{N}_{n = 1} E_{t,n} - \left[ \frac{D_{t,n}}{M^{n}_{V}} * P_{G} \right]
	\end{equation}
	But, since we are estimating the reward for individual vehicles, we get for vehicle $n$ at time step $t$, the average profits becomes:
	\begin{equation}
	\mathbb{P}_{t,n} =  E_{t,n} - \left[ \frac{D_{t,n}}{M^{n}_{V}} * P_{G} \right]
	\end{equation}
	\item Minimize the number of utilized vehicles/resources. We capture this by minimizing the number of vehicles that become active from being inactive at time step $t$. 
	Let $e_{t,n}$ represent whether vehicle $n$ is non-empty at time step $t$. The total number of vehicles that recently became active at time $t$ is given by:
	\begin{equation}
	\vspace{-0.1in}
	e_{t} = \sum^{N}_{n = 1} \left[\text{max}(e_{t,n} - e_{t-1,n}, 0) \right]
	\end{equation}
	Although we are minimizing the number of active vehicles in time step $t$, if the total distance or the total trip time of the passengers increase, it would be beneficial to use an unoccupied vehicle instead of having existing passengers encounter a large undesired delay.
\end{enumerate}
Having defined all our objective terms, we represent the DQN reward function as a weighted sum of these terms as follows:
\vspace{-0.1in}
\begin{equation}\label{reward}
r_{t} = - \left[ \beta_{1} \text{diff}_{t} + \beta_{2} T_{t}^{D} + \beta_{3} \Delta t \right] + \beta_{4} \mathbb{P}_{t} - \beta_{5} e_{t}
\end{equation}

Note, from equation \eqref{total}, that we maximize the discounted reward over a time frame. The negative sign here indicates that we want to minimize the terms within the bracket. 

Note that weights $\beta_{1}, \beta_{2}, \beta_{3}, \beta_{4} \text{ and } \beta_{5}$ depend on the weight factors of each of the objectives. Further, we maximize the discounted reward over a time frame, and the negative sign here indicates that we want to minimize the terms within the function. Finally, note that the reward for vehicle $n$ is 0 if it decides to only serve the  passengers on-board (if, any). Therefore, we focus on the scenario where vehicle $n$ decides to serve a new user and it is willing to take a detour at time $t$. In this case, the reward $r_{t,n}$ for vehicle $n$ at time slot $t$ is represented equation \eqref{individual}, where the objectives above are mapped to: (1) $C_{t,n}$: number of customers served by vehicle $n$ at time $t$, (2), (3) dispatch time and extra travel time are the same, denoted by: $T_{t,n}^{D}$, and $T_{t,n}^{E}$. (4) average profit for vehicle $n$ at time $t$, $\mathbb{P}_{t,n}$. In this case, the reward $r_{t,n}$ for vehicle $n$ at time $t$ is represented by Eq. \eqref{individual} in Section \ref{MP}.


In equation \eqref{individual}, the last term captures the status of vehicle $n$ where $e_{t,n}$ is set to 1 if vehicle $n$ was empty and then becomes occupied at time $t$ (even if by one passenger), however, if it was already occupied and just takes a new customer, $e_{t,n}$ is 0. The intuition here is that if an already occupied vehicle serves a new user, the congestion and fuel costs will be less when compared to when an empty vehicle serves that user. Note that if we make $\beta_{3}$ very large, it will disincentivize passengers and drivers from making detours to serve other passengers, Thus, the setting becomes similar to in \cite{fleet_oda}, with no carpooling.

In our algorithm, we use reinforcement learning to learn the reward function stated in (Eq. \eqref{individual}) using DQN. Through learning the probabilistic dependence between the action and the reward function, we learn the Q-values associated with the probabilities $P(r_{t}\mid a_{t},s_{t})$ over time by feeding the current states of the system. Instead of assuming any specific structure, our model-free approach learns the Q-values dynamically using convolutional neural networks whose architecture is described in Appendix \ref{DQN_Arch} The Q-values are then used to decide on the best dispatching action to take for each individual vehicle. Since the state space is large, we don't use the full representation of $s_{t}$, instead a map-based input is used to alleviate this massive computing.

\begin{figure}
    \centering
    \includegraphics[width=0.37\textwidth]{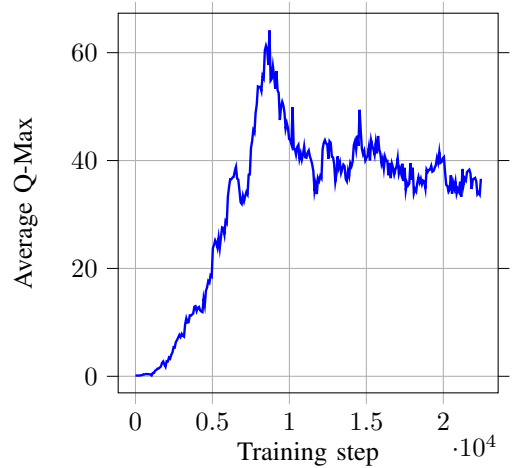}
    \vspace{-0.1in}
    \caption{Convergence of Q-values during training (at approximately 15k training steps.}\label{fig:Learning Curve}
    \vspace{-0.1in}
  \end{figure}

\vspace{-0.1in}
\subsection{Learning Expected Discounted Reward (Q-values)} \label{learn}
In our algorithm, deep queue networks are utilized to dynamically generate optimized values.  Fig. \ref{fig:Learning Curve} shows the convrgence of average Q-max during training. This technique of learning is characterized by its high adaptability to dynamic features in the system, which is why it is widely adopted in modern decision-making tasks. The optimal action-value function for vehicle $n$ is defined as the maximum expected achievable reward. Thus, for any policy $\pi_{t}$ we have:
\begin{equation}
\begin{multlined}
Q^{*}(s,a)\; = \; max_{\pi}\; \mathbb{E} \; \left[\;\sum^{\infty}_{k=t} \eta^{k-t} r_{k,n} \right.  \\  \mid (s_{t,n} = s, a_{t,n} = a, \pi_{t})\; \left. \right]
\end{multlined}
\end{equation}
where $0 < \eta < 1$ is the discount factor for the future. If $\eta$ is small (large, resp.), the dispatcher is more likely to maximize the immediate (future, resp.) reward. At any time slot $t$, the dispatcher monitors the current state $s_{t}$ and then feeds it to the neural network (NN) to generate an action. In our algorithm, we utilize a neural network to approximate the Q function in order to find the expectation.

For each vehicle $n$, an action is taken such that the output of the neural network is maximized. The learning starts with no knowledge and actions are chosen using a greedy scheme by following the Epsilon-Greedy method. Under this policy, the
agent chooses the action that results in the highest Q-value with probability $1-\epsilon$, otherwise, it selects a random action. The $\epsilon$ reduces linearly from 1 to 0.1 over $T_{n}$ steps. For the $n^{th}$ vehicle, after choosing the action and according to the reward $r_{t,n}$, the Q-value is updated with a learning factor $\sigma$ as follows:
\vspace{-0.05in}
\begin{equation}
\begin{multlined}
Q^{'}(s_{t,n},a_{t,n})\; \leftarrow \; (1 - \sigma) \; Q(s_{t,n},a_{t,n}) \; \\ + \;
\sigma \; [r_{t,n} + \eta \; \text{max}_{a} \; Q(s_{t+1,n},a)\; ]
\end{multlined}
\end{equation}
Similar to $\epsilon$, the learning rate $\sigma$ is also reduced linearly from 0.1 to 0.001 over 10000 steps. We note that an artificial neural network is needed to maintain a large system space. When updating these values, a loss function $L_{i}(\theta_{i})$ is used to compute the difference between the predicted Q-values and
the target Q-values, i.e.,
\begin{equation}
\begin{multlined}
L_{i}(\theta_{i}) \; = \; \mathbb{E}  \; \left[\; \left( (r_{t} \; + \; \eta \; \text{max}_{a} Q(s,a; \bar{\theta_{i}})) \right. \right. \\ - Q(s,a; \theta_{i}) \left. \right)^{2} \; \left. \right]
\end{multlined}
\end{equation}
where $\theta_{i}, \bar{\theta_{i}}$, are the weights of the neural networks. This
above expression represents the mean-squared error in the Bellman equation where the optimal values are approximated with a target value of $r_{t} \; + \; \eta \; \text{max}_{a} Q(s,a; \bar{\theta_{i}})$, using the weight $\bar{\theta_{i}}$ from some previous iterations.

\onecolumn
\section{Enlarged Figures}
\begin{figure*}[th]
\captionsetup{justification=centering, font=small, format=hang}
\centering
\includegraphics[trim= 100 20 20 20,  width=0.8\textwidth]{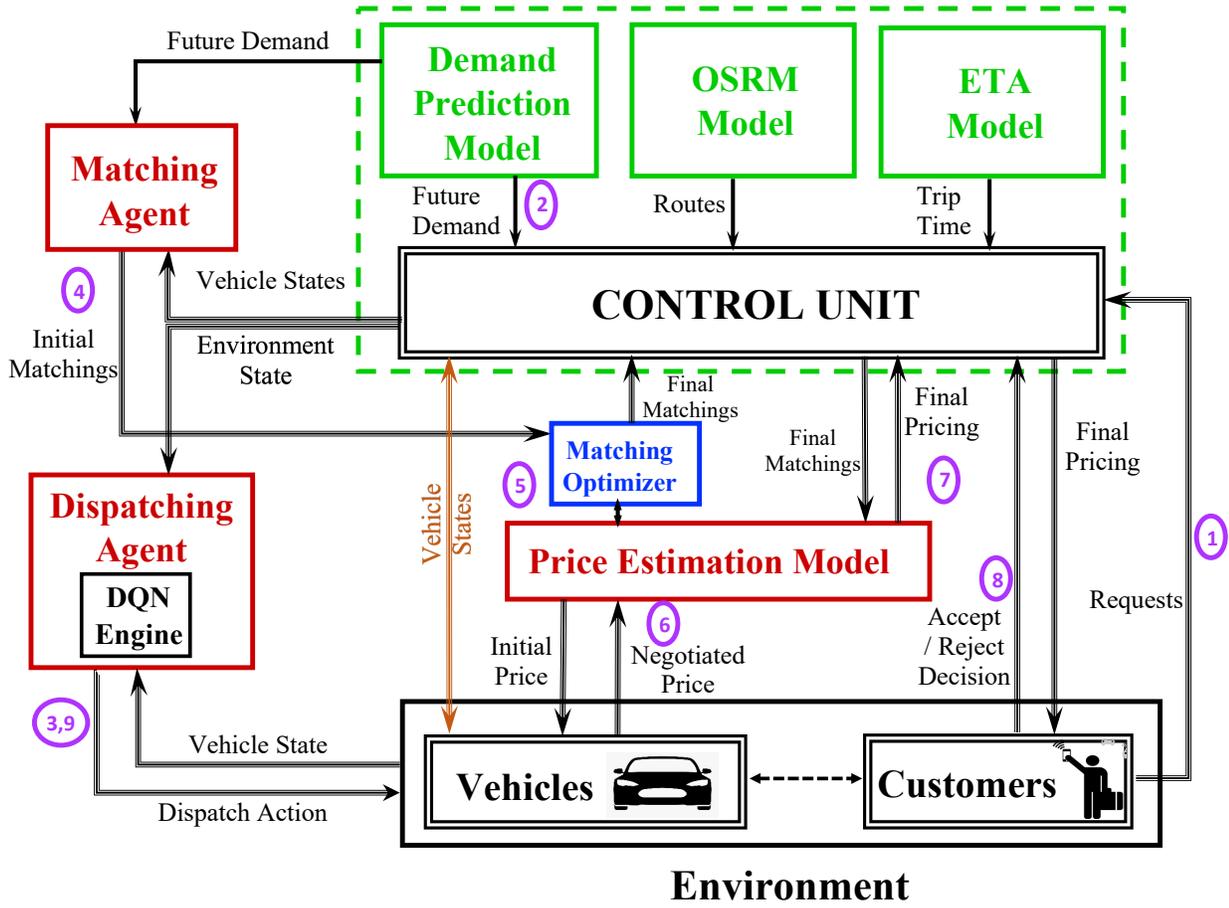}
\caption{Overall architecture of the proposed framework} \label{large}
\end{figure*}

\begin{figure*}[h!]
  \centering
 \begin{tabular}{c}
 \begin{subfigure}{0.46\textwidth}
\includegraphics[trim=10 0 40 25,width=\textwidth]{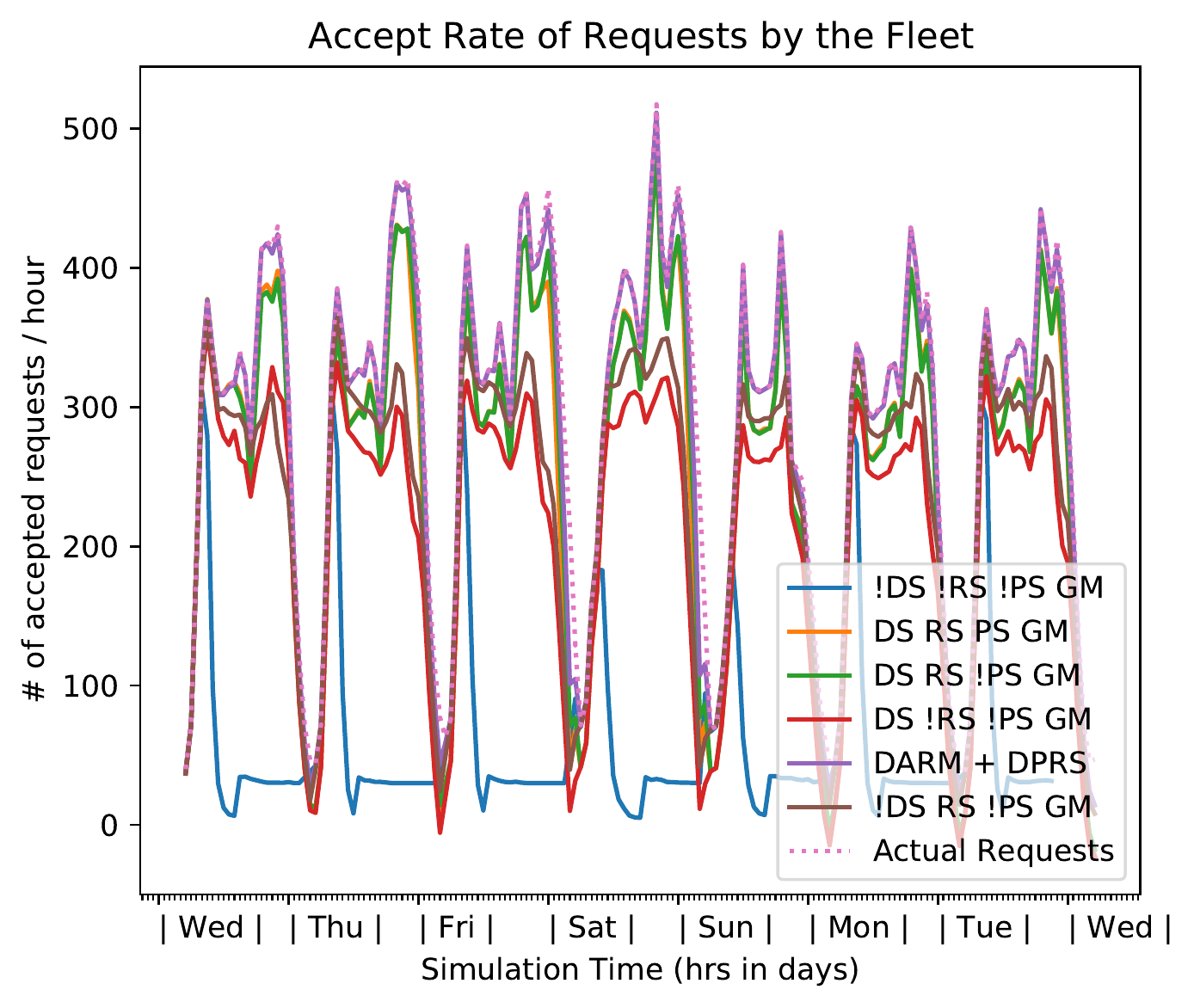} 
\end{subfigure} \\ [1.7em]
\begin{subfigure}{0.49\textwidth}
\includegraphics[trim=10 0 50 0,width=\textwidth]{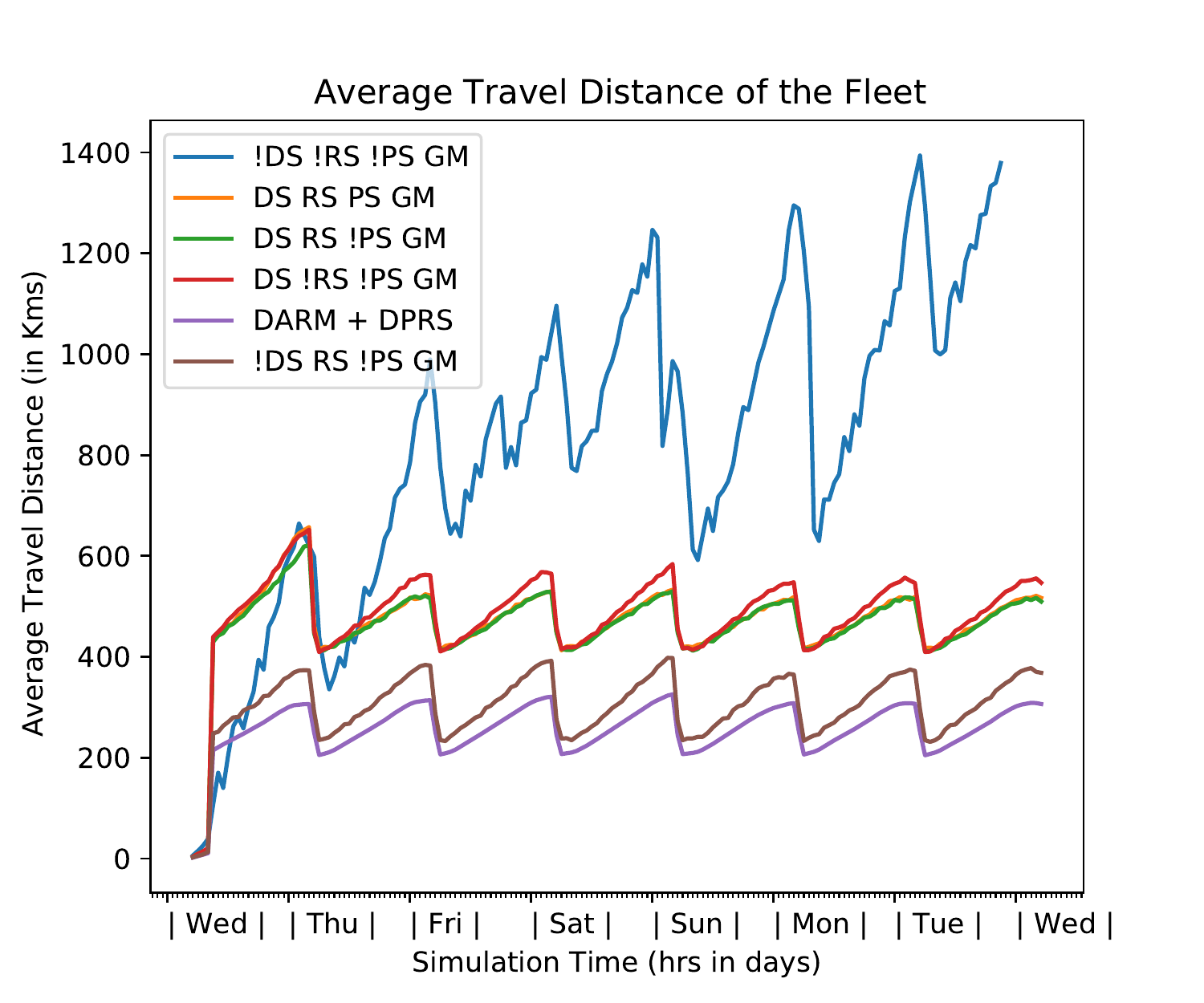} 
\end{subfigure}  \\ [1.7em]
\begin{subfigure}{0.49\textwidth}
\includegraphics[trim=10 0 50 0, width=\textwidth]{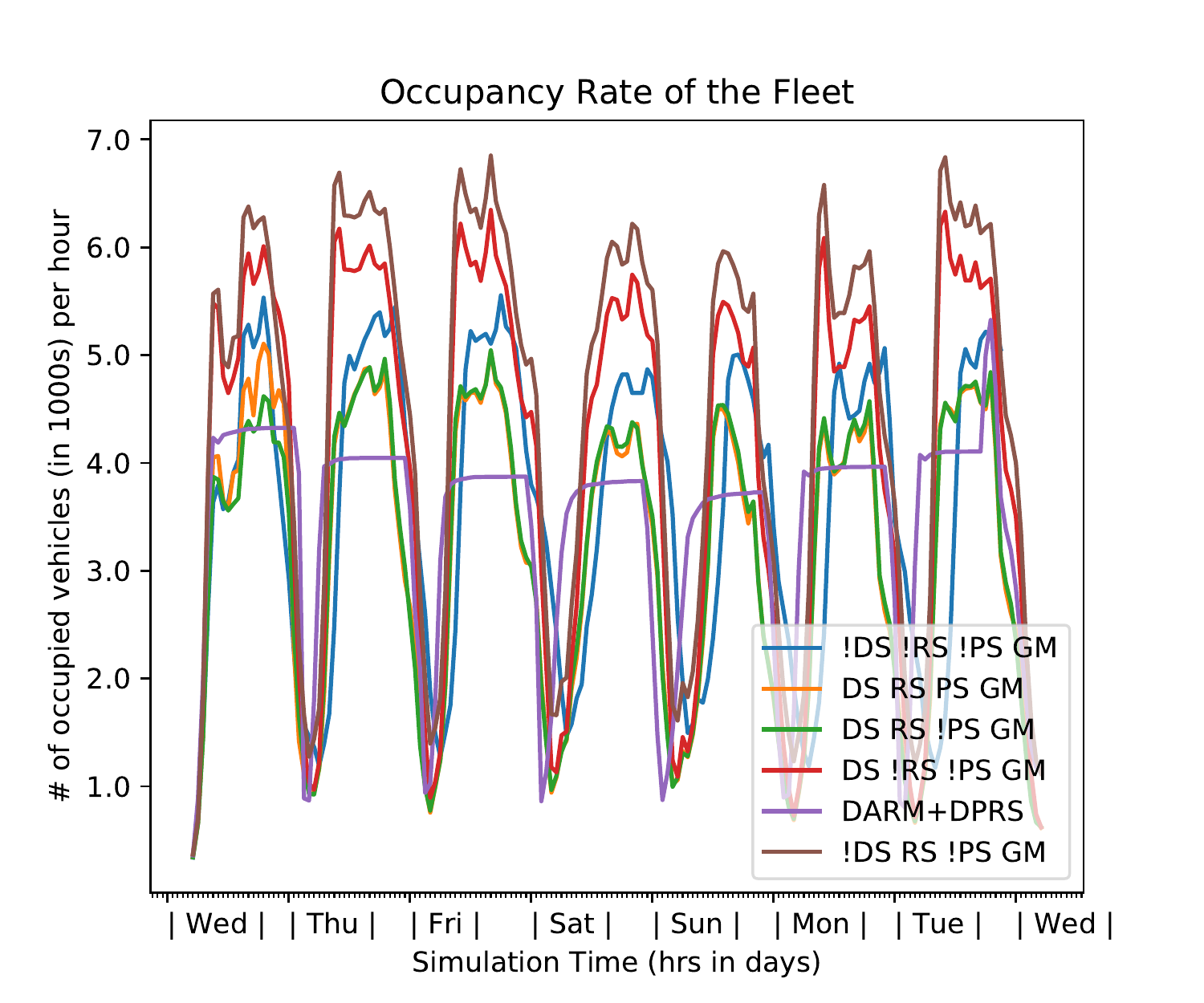} 
\end{subfigure}
\end{tabular}
\caption{Performance Metrics of the proposed algorithm and the  baselines} \label{big}
\end{figure*}

\begin{figure*}[h!]
  \centering
 \begin{tabular}{c}
 \begin{subfigure}{0.8\textwidth}
\includegraphics[trim=30 10 30 5, width=\textwidth]{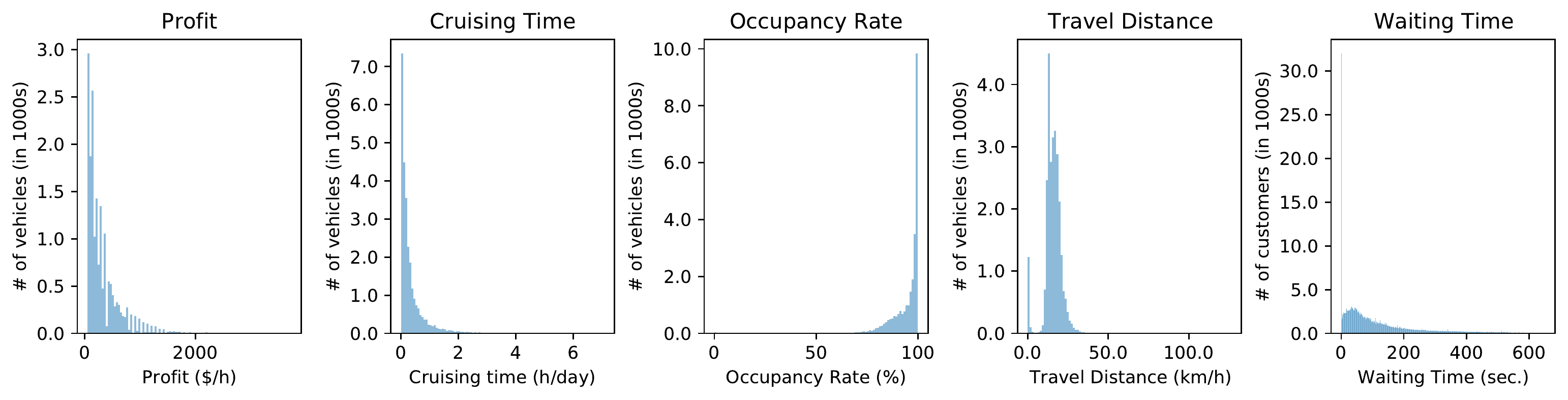} 
\caption{\small Performance Metrics of our Joint Framework \\ ``DARM + DPRS" }
\end{subfigure} \\ [1.2em]
\begin{subfigure}{0.8\textwidth}
\includegraphics[trim=30 10 30 5, width=\textwidth]{DPRS_Metrics.pdf} 
\caption{\small Performance Metrics of (D, RS, PS, GM) Baseline \\ ``DPRS with Greedy Matching"}
\end{subfigure}  \\ [1.2em]
\begin{subfigure}{0.8\textwidth}
\includegraphics[trim=30 10 30 5, width=\textwidth]{DP_Metrics.pdf} 
\caption{\small Performance Metrics of (D, RS, !PS, GM) Baseline \\ 	``Deep\_Pool in \cite{deep_pool}"}
\end{subfigure}  \\ [1.2em]
\begin{subfigure}{0.8\textwidth}
\includegraphics[trim=30 10 30 5, width=\textwidth]{ODA_Metrics.pdf} 
\caption{\small  Performance Metrics of (D, !RS, !PS, GM) Baseline \\ ``Dynamic Taxi Fleet Management in \cite{fleet_oda}" }
\end{subfigure}
\end{tabular}
\caption{Histograms of Performance Metrics for the Proposed Algorithm and the Baselines} \label{enlarged}
\end{figure*}
\end{document}